\documentclass[letterpaper,titlepage,11pt]{article}
\usepackage{hyperref}

\usepackage{amssymb,amsmath,amsfonts}
\usepackage{epsfig}

\def\clock{{\count0=\time
           \divide\count0 60
           \ifnum\count0<10 0\fi\the\count0
           \multiply\count0 -60 \advance\count0 \time
           :\ifnum\count0<10 0\fi \the\count0
         }}

\newcommand{\timestamp}{{\small\vbox{\hbox{\tt\jobname.tex}
\hbox{\the\day/\the\month/\the\year, \clock}}}}

\newcommand{\CA}{\mathcal{A}}

\newcommand{\CH}{\mathcal{H}}

\newcommand{\CO}{\mathcal{O}}

\newcommand{\CN}{\mathcal{N}}

\newcommand{\CV}{\mathcal{V}}

\newcommand{\Z}{\mathbb{Z}}

\newcommand{\R}{\mathbb{R}}

\newcommand{\nn}{\nonumber}

\newcommand{\spa}{\,, \ \ }

\newcommand{\tr}{\mathop{{\rm Tr}}}

\newcommand{\ads}{\mbox{AdS}}
\newcommand{\gym}{g_{\rm YM}}

\newcommand{\bchi}{\bar{\chi}}

\newcommand{\Li}{\textrm{Li}}

\numberwithin{equation}{section}

\begin{document}

\begin{titlepage}

\rightline{\vbox{\small\hbox{\tt NORDITA-2007-21} }}
 \vskip 2 cm

\centerline{\LARGE \bf Decoupling limits of $\CN=4$ super Yang-Mills
on $\R \times S^3$} \vskip 1.8cm

\centerline{\large {\bf Troels Harmark$\,^{1}$}, {\bf Kristj{\'a}n R.\ Kristj{\'a}nsson$\,^{2}$} and {\bf Marta Orselli$\,^{1}$} }

\vskip 0.6cm

\begin{center}
\sl $^1$ The Niels Bohr Institute  \\
\sl  Blegdamsvej 17, 2100 Copenhagen \O , Denmark \\
\vskip 0.4cm
\sl $^2$ NORDITA \\
\sl
Roslagstullsbacken 23,
10691 Stockholm, Sweden \\
\end{center}
\vskip 0.6cm

\centerline{\small\tt harmark@nbi.dk, kristk@nordita.org, orselli@nbi.dk}

\vskip 1.5cm

\centerline{\bf Abstract} \vskip 0.2cm \noindent We find new
decoupling limits of $\CN=4$ super Yang-Mills (SYM) on $\R \times
S^3$ with gauge group $SU(N)$. These decoupling limits lead to
decoupled theories that are much simpler than the full $\CN=4$ SYM
but still contain many of its interesting features. The decoupling
limits correspond to being in a near-critical region, near a point
with zero temperature and critical chemical potentials. The new
decoupling limits are found by generalizing the limits of
hep-th/0605234 to include not only the chemical potentials for the
$SU(4)$ R-symmetry of $\CN=4$ SYM but also the chemical potentials
corresponding to the $SO(4)$ symmetry. In the decoupled theories it
is possible to take a strong coupling limit in a controllable manner
since the full effective Hamiltonian is known. For planar $\CN=4$ SYM
on $\R \times S^3$ all the decoupled theories correspond to fully
integrable spin chains. We study the thermodynamics of the decoupled
theories and find the Hagedorn temperature for small and large
values of the effective coupling. We find an alternative formulation
of the decoupling limits in the microcanonical ensemble. This leads
to a characterization of certain regimes of weakly coupled $\CN=4$
SYM in which there are string-like states. Finally, we find a
similar decoupling limit for pure Yang-Mills theory, which for the
planar limit leads to a fully integrable decoupled theory.


\end{titlepage}

\pagestyle{plain} \setcounter{page}{1}

\tableofcontents


\section{Introduction}
\label{sec:intro}

The AdS/CFT correspondence conjectures a precise duality between
$\CN=4$ supersymmetric Yang-Mills (SYM) theory and type IIB string theory on
$\ads_5\times S^5$
\cite{Maldacena:1997re,Gubser:2002tv,Witten:1998qj}. As a
consequence of this correspondence, it is believed that weakly
coupled string theory on $\ads_5\times S^5$ emerges from large $N$
$\CN=4$ SYM with gauge group $SU(N)$ in the limit of large 't Hooft
coupling. This is in accordance with the ideas of 't Hooft of the
emergence of string theory from gauge theory when the number of
colors is sent to infinity \cite{'tHooft:1974jz}.

However, taking the strong 't Hooft coupling limit of large $N$
$SU(N)$ $\CN=4$ SYM is a highly non-trivial task. For planar $\CN=4$
SYM, significant progress has been made, in particular with the idea
of integrable spin chains as being the connecting link between gauge
theory and string theory \cite{Minahan:2002ve}. However, despite the
remarkable progress, it seems a highly difficult task to use this to
understand $\CN=4$ SYM beyond the planar diagrams, and it is
furthermore difficult to generalize the methods to other gauge
theories.

In this paper, we take a different route, following the papers
\cite{Harmark:2006di,Harmark:2006ta,Harmark:2006ie,Harmark:2007et}.
The idea is to consider decoupling limits of $\CN=4$ SYM on $\R
\times S^3$ with gauge group $SU(N)$. By taking such a decoupling
limit, the remaining decoupled theory is significantly simpler than
the full $\CN=4$ SYM theory and this makes it possible to take a
strong coupling limit of the decoupled theory in a controllable
manner.

The decoupling limits are taken by considering the partition
function in the grand canonical ensemble, which depends on the
temperature and the chemical potentials. The chemical potentials are
$\omega_1$ and $\omega_2$, corresponding to the two charges $S_1$
and $S_2$ of the $SO(4)$ group of $S^3$, and $\Omega_1$, $\Omega_2$
and $\Omega_3$ corresponding to the three R-charges $J_1$, $J_2$ and
$J_3$. The idea is to consider the behavior of $\CN=4$ SYM on $\R
\times S^3$ near a critical point of zero temperature and critical
chemical potential $(\omega_1,\omega_2,\Omega_1,\Omega_2,\Omega_3) =
(n_1,n_2,n_3,n_4,n_5)$, with $n_i$ being fixed numbers. Writing then
$(\omega_1,\omega_2,\Omega_1,\Omega_2,\Omega_3) = (n_1 \Omega,n_2
\Omega,n_3 \Omega,n_4 \Omega,n_5 \Omega)$, with $\Omega$ a parameter
ranging from 0 to 1, the decoupling limits of $\CN=4$ SYM on $\R
\times S^3$ take the form
\begin{equation}
\label{DECLIM}
 \Omega \rightarrow 1,\quad
 \tilde{T} \equiv \frac{T}{1-\Omega} \ \mbox{fixed},\quad
 \tilde{\lambda} \equiv \frac{\lambda}{1-\Omega} \ \mbox{fixed},\quad
 N \ \mbox{fixed},
\end{equation}
where $\lambda$ is the 't Hooft coupling.
In such a decoupling limit we show that the full partition function
of $\CN=4$ SYM on $\R \times S^3$ reduces to ($\tilde{\beta} =
1/\tilde{T}$)
\begin{equation}
\label{ZZZ}
Z(\tilde{\beta}) = \tr \left( e^{-\tilde{\beta} ( D_0 + \tilde{\lambda} D_2 ) } \right) ,
\end{equation}
where the trace runs over a subset of the states, and the $D_0$ and
$D_2$ operators come from the weak coupling expansion of the
dilatation operator $D = D_0 + \lambda D_2 + \CO( \lambda^{3/2} )$,
with $D_0$ being the bare scaling dimension and $D_2$ the one-loop
contribution. The trace in Eq.~\eqref{ZZZ} runs over the subset of
states corresponding to the set of gauge-invariant operators of
$\CN=4$ SYM fulfilling the equation $D_0 = J$, with $J \equiv n_1
S_1 + n_2 S_2 + n_3 J_1 + n_4 J_2 + n_5 J_3$.

The general idea with these decoupling limits is then that the full
$\CN=4$ SYM reduces to a subsector, and that the full effective
Hamiltonian reduces to the truncated Hamiltonian $D_0 +
\tilde{\lambda} D_2$ containing only the zero and one-loop terms.
This makes it possible to take the large $\tilde{\lambda}$ limit.
Since for an expansion for small $\tilde{\lambda}$ a contribution at
order $\tilde{\lambda}^n$ origins from a $\lambda^n$ term in the
full theory we can in this sense say that $\tilde{\lambda}
\rightarrow \infty$ corresponds to taking a strong coupling limit of
the theory, even though $\lambda$ is small in the limit
\eqref{DECLIM}. Therefore, we are able to take explicitly a strong
coupling limit by selecting only a subclass of the diagrams for the
full theory.

A particular limit of the above kind was found and studied in
\cite{Harmark:2006di,Harmark:2006ta} with the critical point given by
$(n_1,n_2,n_3,n_4,n_5)=(0,0,1,1,0)$. In the limit \eqref{DECLIM} all
the states decouple except for those in the $SU(2)$ sector. For
the single-trace operators of planar $\CN=4$ SYM, the $\tilde{\lambda}D_2$
term corresponds to the Hamiltonian of the ferromagnetic
$XXX_{1/2}$ Heisenberg spin chain. Therefore, weakly coupled planar
$\CN=4$ SYM becomes equivalent to the Heisenberg spin chain in this
decoupling limit. In \cite{Harmark:2006ta} this was used to find the
spectrum in the limit of large $\tilde{\lambda}$. The spectrum for
$\tilde{\lambda} \rightarrow \infty$ was shown to be given by the
spectrum of free magnons in the Heisenberg spin chain.

The AdS/CFT correspondence states that planar $\CN=4$ SYM on $\R
\times S^3$ is dual to tree-level type IIB string theory on
$\ads_5\times S^5$. Thus, the decoupling limit \eqref{DECLIM} with
$(n_1,n_2,n_3,n_4,n_5)=(0,0,1,1,0)$ of planar $\CN=4$ SYM on $\R
\times S^3$ is dual to the corresponding decoupling limit of
tree-level string theory on $\ads_5\times S^5$ \cite{Harmark:2006ta}. By
employing a certain Penrose limit \cite{Bertolini:2002nr}, we found in \cite{Harmark:2006ta} the
spectrum for large $\tilde{\lambda}$ and matched this to the
spectrum found on the gauge theory side, for large $J = J_1+J_2$. We
furthermore used this to match the Hagedorn temperature as computed
on the gauge theory and string theory sides. The match of the
spectrum and the Hagedorn temperature means that the strong coupling
limit $\tilde{\lambda} \rightarrow \infty$ on the gauge theory side
correctly matches the same decoupled regime in string theory.
Therefore, the decoupling limit \eqref{DECLIM} provides us with a
precise way to match gauge theory with string theory.

In this paper we find all the decoupling limits of the form
\eqref{DECLIM}, where
$(\omega_1,\omega_2,\Omega_1,\Omega_2,\Omega_3)
=(n_1,n_2,n_3,n_4,n_5)$ corresponds to a critical value for the
chemical potentials of $\CN=4$ SYM on $\R \times S^3$. We find a
total of fourteen such decoupling limits of $\CN=4$ SYM on $\R
\times S^3$, three of them found previously in
\cite{Harmark:2006di}. These fourteen limits correspond to fourteen
different subgroups of the total symmetry group $PSU(2,2|4)$ of
$\CN=4$ SYM. We show that in the planar limit, each of the fourteen
decoupled theories corresponds to a fully integrable spin chain
(previously considered in \cite{Beisert:2004ry}). Some of these
decoupled theories are well-known theories in the Condensed Matter
literature, thus in this sense we have found limits of $\CN=4$ SYM
on $\R \times S^3$ where it reduces to known Condensed Matter
theories. However, when going beyond the planar part of $\CN=4$ SYM,
the decoupling limits give rise to new decoupled theories.

Of the fourteen decoupling limits that we find, two give rise to
trivial decoupled theories. The remaining twelve non-trivial
decoupled theories are divided into nine theories with scalars and
three without scalars.  We explain that the presence of scalars is
crucial for how the theory behaves in the large $\tilde{\lambda}$
limit. One of the theories with scalars has a $SU(1,2|3)$ symmetry,
and we show that all the other decoupled theories can be seen to be
a subsector of the theory with $SU(1,2|3)$ symmetry.

We consider in detail the decoupled theories in the planar limit. We employ
recent results in the literature to write down the Bethe equations for the
decoupled theories, and use this to find the low energy limit of the
spectrum for each theory.

We analyze furthermore the thermodynamics of the decoupled theories
in the planar limit. For each theory we compute the partition
function and the Hagedorn temperature for zero coupling, and for
small $\tilde{\lambda}$ we find the first correction in
$\tilde{\lambda}$. For the nine theories with scalars we use the
results for the low energy spectra to determine the Hagedorn
temperature for large $\tilde{\lambda}$. We furthermore explain why
the large $\tilde{\lambda}$ behavior for the three theories without
scalars is difficult to attain.

We provide an equivalent formulation of the decoupling limits
\eqref{DECLIM} in the microcanonical ensemble, $i.e.$ with the
limits formulated in terms of $D$, $S_1$, $S_2$, $J_1$, $J_2$ and
$J_3$. This is crucial for translating the limits to the string side
of the AdS/CFT correspondence, but it is also highly important in
order to understand precisely which regimes of $\CN=4$ SYM on $\R
\times S^3$ the decoupled theories correspond to. It is furthermore
a check that the decoupling limits are consistent. We find in
particular that for the nine non-trivial theories with scalars, the
states in $\CN=4$ SYM on $\R \times S^3$ that dominate in the strong
coupling limit $\tilde{\lambda} \rightarrow \infty$ are the ones in
the regime
\begin{equation}
|D-J| \ll \lambda \ll 1, \quad J \gg 1
\end{equation}
with $J \equiv n_1 S_1 + n_2 S_2 + n_3 J_1 + n_4 J_2 + n_5 J_3$.

As we discuss in the paper, formulating the limits in the
microcanonical ensemble also means that we can think of the limits
as being taken of the gauge-invariant operators of $\CN=4$ SYM on
$\R^4$, rather than of the states of $\CN=4$ SYM on $\R \times S^3$.

Finally, we use our insights obtained for $\CN=4$ SYM on $\R \times
S^3$ to formulate a new decoupling limit of pure Yang-Mills (YM) theory on
$\R \times S^3$. Our new decoupling limit of pure YM shares many
features with one of the decoupling limits for $\CN=4$ SYM,
corresponding to $(n_1,n_2,n_3,n_4,n_5)=(1,1,0,0,0)$. For planar
pure YM we show that the decoupled theory obtained from the
decoupling limit corresponds to an integrable spin chain. We furthermore
analyze the large $\tilde{\lambda}$ limit and discuss the
implications for finding a string-dual of pure YM.

\section{New decoupling limits}
\label{sec:newlim}

In this section we generalize the recently found decoupling limits
\cite{Harmark:2006di} for weakly coupled $\CN=4$ SYM on $\R \times
S^3$ with gauge group $SU(N)$ to include chemical potentials for the
R-charges of the $SU(4)$ R-symmetry as well as the Cartan generators
of the $SO(4)$ symmetry group of $S^3$. The limits are taken of the
thermal partition function of $\CN=4$ SYM on $\R \times S^3$ in the
grand canonical ensemble and they are valid for finite $N$. For each
decoupling limit, only a subset of the states of $\CN=4$ SYM on $\R
\times S^3$ survive and the effective Hamiltonian truncates to
include only the tree-level and one-loop terms of the full theory.
In Section \ref{sec:declist} we list all of the fourteen different
decoupling limits that one can have, along with the field content
and the symmetry algebra for each of the decoupled theories.
Finally, we show in Section~\ref{sec:closeD2} that all the decoupled
sectors are closed under the action of the one-loop dilatation
operator $D_2$.

\subsection{General considerations}
\label{sec:gencon}

We consider $\CN=4$ SYM on $\R \times S^3$ with gauge group $SU(N)$. We define the 't Hooft coupling as%
\footnote{The $4\pi^2$ factor is included in the 't Hooft coupling
for our convenience.}
\begin{equation}
\lambda = \frac{\gym^2 N}{4\pi^2},
\end{equation}
where $\gym$ is the Yang-Mills coupling of $\CN=4$ SYM. For $\CN=4$
SYM on $\R \times S^3$ the states are mapped to the operators of
$\CN=4$ SYM on $\R^4$, with the energy of a state mapped to the
scaling dimension of the operator (we assume here that the radius of
$S^3$ equals one).  Since we are on an $S^3$ we only have gauge
singlet states. This means that the set of operators $M$ that we
should consider is the set of gauge invariant operators, which are
all the possible linear combinations of the multi-trace operators
\begin{equation}
\tr \left( A^{(1)}_1 A^{(1)}_2 \cdots A^{(1)}_{L_1}  \right) \tr \left( A^{(2)}_1
A^{(2)}_2 \cdots A^{(2)}_{L_2}  \right) \cdots \tr \left( A^{(k)}_1 A^{(k)}_2
\cdots A^{(k)}_{L_k}  \right).
\end{equation}
Here $A^{(i)}_j \in \CA$, with $\CA$ being the set of letters which
is the singleton representation of $psu(2,2|4)$. We review the set
of letters $\CA$ in detail in Section~\ref{sec:systexpl}.  Each
state carries quantum numbers according to the Cartan generators of
$psu(2,2|4)$. These are the energy $E$, the two angular momenta
$S_1,S_2$ corresponding to the $SO(4)$ symmetry of $S^3$, and the
three R-symmetry charges $J_1$, $J_2$, $J_3$ corresponding to the
Cartan generators of the $SU(4)$ R-symmetry subgroup of
$PSU(2,2|4)$. For the corresponding operator we have the scaling
dimension $D$, along with angular momenta $S_1,S_2$ and the
R-symmetry charges $J_1,J_2,J_3$.

In general, we can write the partition function of $\CN=4$ SYM on
$\R \times S^3$ with
gauge group $SU(N)$ in the grand canonical ensemble as
\begin{equation}
\label{genZ}
Z_{\lambda,N}(\beta,\omega_1,\omega_2,\Omega_1,\Omega_2,\Omega_3)
= {\tr}_M \left[ \exp \left( - \beta D + \beta \sum_{a=1}^2
\omega_a S_a + \beta \sum_{i=1}^3 \Omega_i J_i \right) \right]
\end{equation}
where $T=1/\beta$ is the temperature, $\omega_1,\omega_2$ are
the chemical potentials corresponding to $S_1,S_2$, and
$\Omega_1,\Omega_2,\Omega_3$ are the chemical potentials
corresponding to $J_1,J_2,J_3$. $M$ is the set of gauge invariant
operators defined above. Note that the dependence on $\lambda$ enters only through the dilatation operator $D$, while the $N$ dependence enters through $D$ and the set of operators $M$.

In the following we are interested in the situation in which some or
all of the chemical potentials are set to be proportional to the
same parameter $\Omega$ and the rest are zero. We write this in
general as
\begin{equation}
\label{omegas} (\omega_1,\omega_2,\Omega_1,\Omega_2,\Omega_3) =
(n_1\Omega,n_2\Omega,n_3\Omega,n_4\Omega,n_5\Omega)
\end{equation}
with $n_i$ being real numbers. The parameter $\Omega$ is ranging from 0 to 1. As we shall see below, the numbers $(n_1,n_2,n_3,n_4,n_5)$ correspond to critical values of the set of chemical potentials $(\omega_1,\omega_2,\Omega_1,\Omega_2,\Omega_3)$. Thus, as $\Omega$ is sent towards 1, we approach a critical value of the set of chemical potentials.

Employing \eqref{omegas}, we can then write the partition
function \eqref{genZ} as
\begin{equation}
\label{ompart}
Z_{\lambda,N}(\beta,\Omega) = {\tr}_M \left[ e^{ - \beta D + \beta \Omega J }
\right] = {\tr}_M \left[ e^{ - \beta ( D - J) - \beta (1-\Omega) J }
\right]
\end{equation}
where we defined
\begin{equation}
\label{defJ} J \equiv n_1 S_1 + n_2 S_2 + n_3 J_1 + n_4 J_2 + n_5
J_3 .
\end{equation}

In general we can write the dilatation operator $D$ for small $\lambda$ as
\begin{equation}
\label{dilop} D = D_0 + \lambda D_2 + \lambda^{3/2} D_3 + \lambda^2
D_4 + \cdots
\end{equation}
Here $D_0$ corresponds to the bare scaling dimension,
$D_2$ the one-loop correction, and so on.

We now want to consider taking a limit with the temperature
$T=1/\beta$ going to zero.
Focusing first on the free case $\lambda=0$ the partition function is
\begin{equation}
\label{freep}
Z_{\lambda=0,N}(\beta,\Omega)  =
{\tr}_M \left[ e^{ - \beta ( D_0 - J) - \beta (1-\Omega) J } \right].
\end{equation}
For all the letters in $\CA$ we have that
$D_0$, $S_1$, $S_2$, $J_1$, $J_2$ and $J_3$ are integers or half-integers.
Thus, given the numbers $(n_1,n_2,n_3,n_4,n_5)$,
we can find a number $b>0$ such that for any state with $D_0-J \neq 0$ we
have that $|D_0-J| \geq b$. Therefore, for $\beta \rightarrow \infty$, all states
with $D_0-J > 0$ decouple from the partition function.
We also see that if we have states with $D_0 - J < 0$ the partition function diverges.
Thus, we restrict ourselves to choices of $(n_1,n_2,n_3,n_4,n_5)$ for which all states obey that $D_0 \geq J$. On the other hand, to avoid that all states decouple
for $\beta \rightarrow \infty$ we see that we need
to choose $(n_1,n_2,n_3,n_4,n_5)$ such that there are states with $D_0 = J$. Considering again
\eqref{freep} we see that to get a non-trivial partition function we need to
keep $\beta (1-\Omega)$ fixed as $\beta\rightarrow \infty$. Thus, taking the limit
\begin{equation}
\label{freelim}
\beta \rightarrow \infty,\quad
\tilde{\beta} \equiv \beta (1-\Omega) \ \mbox{fixed}
\end{equation}
the partition function \eqref{freep} becomes
\begin{equation}
\label{limfreep}
Z_{N}(\tilde{\beta})  = {\tr}_\CH \left[ e^{ - \tilde{\beta} D_0 }
\right]
\end{equation}
where the trace is over the subset $\CH$ of $M$ given by
\begin{equation}
\label{defCH}
\CH = \{ \alpha \in M | (D_0 - J) \alpha = 0 \} .
\end{equation}
We see from \eqref{limfreep} that it makes sense to interpret
$\tilde{T}=1/\tilde{\beta}$ as a temperature for the effective theory
that one gets after taking the decoupling limit \eqref{freelim}.

Considering now the case with non-zero coupling $\lambda$ we see that for small $\lambda$
the partition function \eqref{ompart} is
\begin{equation}
\label{ompart2}
Z_{\lambda,N}(\beta,\Omega)  = {\tr}_M \left[ e^{ - \beta ( D_0 - J) -
\beta \lambda D_2 - \beta (1-\Omega) J + \beta \CO ( \lambda^{3/2} ) }
\right].
\end{equation}
Therefore, we get a non-trivial interaction term only if we keep
$\beta \lambda$ fixed in the $\beta \rightarrow \infty$ limit.

We can now formulate the full decoupling limit. First, we assume that the numbers
$(n_1,n_2,n_3,n_4,n_5)$ are given such that
\begin{itemize}
\item We have that $D_0 \geq J$ for any letter in $\CA$.
\item There exist letters in $\CA$ for which $D_0 = J$.
\end{itemize}
These two conditions are equivalent to demanding that $D_0\geq J$
for all states and that there exist states for which $D_0 = J$. With
respect to $(n_1,n_2,n_3,n_4,n_5)$ we can then define the subset
$\CH$ of $M$ as in \eqref{defCH}. We now take the decoupling limit
\begin{equation}
\label{declim}
\beta \rightarrow \infty, \quad
\tilde{\beta} \equiv \beta (1-\Omega) \ \mbox{fixed}, \quad
\tilde{\lambda} \equiv \frac{\lambda}{1-\Omega} \ \mbox{fixed}, \quad
N \ \mbox{fixed}.
\end{equation}
This brings us near to a point with zero temperature, $\Omega=1$ and zero
coupling. From the above considerations we see that the decoupling limit
\eqref{declim} of the full partition function of $\CN=4$ SYM on $\R \times S^3$
with gauge group $SU(N)$ in the grand canonical ensemble becomes
\begin{equation}
\label{limpart}
Z_{\tilde{\lambda},N}(\tilde{\beta})  = {\tr}_\CH \left[ e^{ - \tilde{\beta} ( D_0 + \tilde{\lambda} D_2 ) }
\right].
\end{equation}
This decoupled partition function can be thought of as a partition function
for a decoupled theory, with the set of operators (and corresponding states) of the theory being $\CH$, as defined in \eqref{defCH}, with the effective
temperature being $\tilde{T}=1/\tilde{\beta}$ and with effective Hamiltonian being $D_0+\tilde{\lambda} D_2$.

Several remarks are in order at this point:
\begin{itemize}
\item The higher loop terms in the dilatation operator $D_{n\geq 3}$ become
negligible in the limit \eqref{declim}. Thus, the interaction
truncates so that it only contains the one-loop contribution $D_2$.
\item So far we have not assumed anything about $N$, thus the above decoupling
limit also works for finite $N$. Therefore, the partition function \eqref{limpart}
depends in general on the three parameters $\tilde{\lambda}$, $N$ and $\tilde{\beta}$.
\item Our requirements for the choice of $(n_1,n_2,n_3,n_4,n_5)$ mean that
$(T,\Omega)=(0,1)$ is a critical point, $i.e.$ that $(T,\omega_1,\omega_2,\Omega_1,\Omega_2,\Omega_3)=(0,n_1,n_2,n_3,n_4,n_5)$ is a critical point. This is one of the reasons why the
limit \eqref{declim} yields an interesting decoupled theory.
\end{itemize}

\subsection{Systematic exploration}
\label{sec:systexpl}

We now examine systematically all the possible decoupling limits of
the type \eqref{declim}. To do this, we first describe the set of
letters $\CA$ and then we proceed to consider which choices of
$(n_1,n_2,n_3,n_4,n_5)$ lead to a decoupling limit.

\subsubsection*{Letters of $\CN=4$ SYM}

The set of letters $\CA$ of $\CN=4$ SYM consists of 6 independent
gauge field strength components, 6 complex scalars and 16 complex
fermions, plus the descendants of these that one gets by applying
the 4 components of the covariant derivative. We describe in the
following how the letters transform in multiplets of the $SO(4)$ and
$SU(4)$ subgroups of $PSU(2,2|4)$. The gauge field strength
components transform in the representations $[0,0,0]_{(1,0)}$ and
$[0,0,0]_{(0,1)}$, where $[k,p,q]_{(j_1,j_2)}$ refers to the
$[k,p,q]$ representation of $SU(4)$ and the $(j_1,j_2)$
representation of $SU(2)\times SU(2) = SO(4)$. The gauge field
strength components have bare dimension $D_0 = 2$. We list the
explicit weights for the gauge field strength components in
Table~\ref{tab:gfield}.
\begin{table}[ht]
\begin{center}
\begin{tabular}{|c||c|c|c|c|c|c|}
\hline & $F_+$ & $F_0$ & $F_-$ & $\bar{F}_+$ & $\bar{F}_0$ & $\bar{F}_-$ \\
\hline $SO(4)$ & $(1,-1)$ & $(0,0)$ & $(-1,1)$ & $(1,1)$ & $(0,0)$ & $(-1,-1)$ \\
\hline $SU(4)$ & $(0,0,0)$ & $(0,0,0)$ & $(0,0,0)$ & $(0,0,0)$ & $(0,0,0)$ & $(0,0,0)$ \\
\hline
\end{tabular}
\caption{Gauge field strength components in $\CN=4$ SYM. \label{tab:gfield}}
\end{center}
\end{table}
The 6 complex scalars transform in the $[0,1,0]_{(0,0)}$
representation. They have bare scaling dimension $D_0=1$ and their
weights are listed in Table~\ref{tab:scalars}.
\begin{table}[ht]
\begin{center}
\begin{tabular}{|c||c|c|c|c|c|c|}
\hline & $Z$ & $X$ & $W$ & $\bar{Z}$ & $\bar{X}$ & $\bar{W}$ \\
\hline $SO(4)$ & $(0,0)$ & $(0,0)$ & $(0,0)$ & $(0,0)$ & $(0,0)$ & $(0,0)$ \\
\hline $SU(4)$ & $(1,0,0)$ &$(0,1,0)$ & $(0,0,1)$ & $(-1,0,0)$ & $(0,-1,0)$ & $(0,0,-1)$ \\
\hline
\end{tabular}
\caption{Scalars of $\CN=4$ SYM. \label{tab:scalars}}
\end{center}
\end{table}
There are 16 complex fermion letters corresponding to the components
of the complex fermionic fields $\chi^\alpha_A$,
$\bar{\chi}_{\dot{\alpha}}^A$, $\alpha,\dot{\alpha}=1,2$,
$A=1,2,3,4$. Half of the 16 complex fermions, denoted
$\chi_1,\chi_2,...,\chi_8$, transform in the  $[0,0,1]_{(1/2,0)}$
representation and the other (conjugate) half, denoted
$\bchi_1,\bchi_2,...,\bchi_8$, transform in the $[1,0,0]_{(0,1/2)}$
representation. The fermions have bare scaling dimension $D_0 =
\frac{3}{2}$. We have listed the $SO(4)$ weights of the fermions in
Table~\ref{tab:fermions1}.
\begin{table}[ht]
\begin{center}
\begin{tabular}{|c||c|c|c|c|}
\hline & $\chi_1$, $\chi_3$, $\chi_5$, $\chi_7$ & $\chi_2$, $\chi_4$,
$\chi_6$, $\chi_8$ & $\bchi_1$, $\bchi_3$, $\bchi_5$, $\bchi_7$ & $\bchi_2$, $\bchi_4$, $\bchi_6$, $\bchi_8$  \\
\hline $SO(4)$ &
$(\frac{1}{2},-\frac{1}{2})$ &
$(-\frac{1}{2},\frac{1}{2})$ &
$(\frac{1}{2},\frac{1}{2})$ &
$(-\frac{1}{2},-\frac{1}{2})$ \\
\hline
\end{tabular}
\caption{$SO(4)$ weights for the fermions of $\CN=4$ SYM. \label{tab:fermions1}}
\end{center}
\end{table}
The $SU(4)$ weights for $\chi_1,\chi_2,...,\chi_8$ are listed in
Table \ref{tab:fermions2} while the ones for
$\bchi_1,\bchi_2,...,\bchi_8$ are listed in Table
\ref{tab:fermions3}. Note that the both the $SO(4)$ and the $SU(4)$
representations are non-trivial for the fermions, contrary to the gauge
field strength and the scalars.
\begin{table}[ht]
\begin{center}
\begin{tabular}{|c||c|c|c|c|}
\hline & $\chi_1$, $\chi_2$ & $\chi_3$, $\chi_4$ & $\chi_5$, $\chi_6$ & $\chi_7$, $\chi_8$  \\
\hline $SU(4)$ &
$(\frac{1}{2},\frac{1}{2},\frac{1}{2})$ &
$(\frac{1}{2},-\frac{1}{2},-\frac{1}{2})$ &
$(-\frac{1}{2},\frac{1}{2},-\frac{1}{2})$ &
$(-\frac{1}{2},-\frac{1}{2},\frac{1}{2})$ \\
\hline
\end{tabular}
\caption{$SU(4)$ weights for the $\chi_1,...,\chi_8$ fermions of $\CN=4$ SYM. \label{tab:fermions2}}
\end{center}
\end{table}
\begin{table}[ht]
\begin{center}
\begin{tabular}{|c||c|c|c|c|}
\hline & $\bchi_1$, $\bchi_2$ & $\bchi_3$, $\bchi_4$ & $\bchi_5$, $\bchi_6$ & $\bchi_7$, $\bchi_8$  \\
\hline $SU(4)$ &
$(-\frac{1}{2},-\frac{1}{2},-\frac{1}{2})$ &
$(-\frac{1}{2},\frac{1}{2},\frac{1}{2})$ &
$(\frac{1}{2},-\frac{1}{2},\frac{1}{2})$ &
$(\frac{1}{2},\frac{1}{2},-\frac{1}{2})$ \\
\hline
\end{tabular}
\caption{$SU(4)$ weights for the $\bchi_1,...,\bchi_8$ fermions of $\CN=4$ SYM. \label{tab:fermions3}}
\end{center}
\end{table}
Finally there are the four components of the covariant derivative.
They are not letters by themselves, but by combining any number of
covariant derivations with a gauge field strength component, a
scalar, or a complex fermion, one gets a letter in $\CA$. The
covariant derivative transforms in the representation
$[0,0,0]_{(1/2,1/2)}$. The covariant derivative components
contribute with $D_0=1$ to the bare scaling dimension of a letter.
We have listed the weights in Table \ref{tab:deriv}.
\begin{table}[ht]
\begin{center}
\begin{tabular}{|c||c|c|c|c|c|c|}
\hline & $d_1$ & $d_2$ & $\bar{d}_1$ & $\bar{d}_2$ \\
\hline $SO(4)$ & $(1,0)$ & $(0,1)$ & $(-1,0)$ & $(0,-1)$ \\
\hline $SU(4)$ & $(0,0,0)$ & $(0,0,0)$ & $(0,0,0)$ & $(0,0,0)$ \\
\hline
\end{tabular}
\caption{Derivative operators of $\CN=4$ SYM. \label{tab:deriv}}
\end{center}
\end{table}

In Appendix \ref{app:oscrep} we review the oscillator representation for
the letters $\CA$ which gives an alternative way of representing $\CA$.

\subsubsection*{Determination of the possible limits}

{}From Section \ref{sec:gencon} we have that a decoupling limit is defined by
$n=(n_1,n_2,n_3,n_4,n_5)$. We now examine systematically what are the
possible choices of $n$ leading to a decoupling limit.

We begin by remarking that with respect to the bosons (the scalars,
the gauge field strength components and the derivatives) we can
choose $n_i \geq 0$ without loss of generality. This is not the case
for the fermions, since the representation of $SO(4)$ is linked to
that of $SU(4)$. However, if we allow for one of the $n_i$ to be
negative, we can choose the other four to be positive. We make the
choice that $n_1$, $n_3$, $n_4$ and $n_5$ should be positive, or
zero, whereas we allow $n_2$ to be negative. This is done without
loss of generality.

As described in Section \ref{sec:gencon} we have two constraints on
$(n_1,n_2,n_3,n_4,n_5)$ in order to have a decoupling limit. The
first constraint is that all letters should obey the inequality $D_0
\geq J$ and the second constraint is that there should be at least
one letter for which $D_0 = J$.

Consider the first constraint. For the three scalars $Z$, $X$, $W$
and the three derivatives $d_1$, $d_2$, $\bar{d}_2$ it implies the
inequalities $n_1 \leq 1$, $-1 \leq n_2 \leq 1$, $n_3 \leq 1$, $n_4
\leq 1$ and $n_5 \leq 1$. We now impose the extra assumption that
$n_3 \geq n_4 \geq n_5$ and $n_1 \geq n_2$, without loss of
generality. We get therefore the following constraints on $n_i$
\begin{equation}
\label{nrange} 0 \leq n_1 \leq 1,\quad
-1 \leq n_2 \leq n_1, \quad
0 \leq n_5 \leq n_4 \leq n_3 \leq 1.
\end{equation}
We now turn to the fermions. It is evident that the number of $+1/2$ in the $SO(4)$
and $SU(4)$ weights is either 0, 2 or 4. From this, we see that only the fermions
with four $+1/2$ in the $SO(4)$ and $SU(4)$ weights can give extra constraints on
the $n_i$ beyond \eqref{nrange}.%
\footnote{This is including the possibility of negative $n_2$.}
The fermions with four $+1/2$ are $\chi_1$, $\chi_2$, $\bchi_3$, $\bchi_5$, $\bchi_7$.
Of these five, only $\chi_1$ and $\bchi_7$ are seen to give new constraints
beyond \eqref{nrange}. Thus, the constraints on the $n_i$ that we get
from the fermions are summarized into the single constraint
\begin{equation}
\label{nfercon}
n_1+n_3+n_4+|n_5-n_2| \leq 3.
\end{equation}
In conclusion we have, with the choices for the $n_i$ made above,
that the constraint that $D_0 \geq J$ for all letters in $\CA$ is
equivalent to the constraints \eqref{nrange} and \eqref{nfercon} for
the $n_i$.

We now turn to the second constraint on the $n_i$ stating that there
should be at least one letter in $\CA$ such that $D_0=J$. Concerning
the complex scalars, it is clear that the number of scalars after
the decoupling, $i.e.$  with $D_0 = J$, is equal to how many of
$n_3$, $n_4$ and $n_5$ are equal to 1. On the other hand, it is
clear that the letters $\bar{Z}$, $\bar{X}$ and $\bar{W}$ can never
be present. For the components of the covariant derivative we have
similarly that the number of derivatives is equal to how many of
$n_1$ and $|n_2|$ are equal to 1, and that the derivative operator
$\bar{d}_1$ cannot be part of any decoupled theory. For the field
strength components, we see that the only two possibilities for a
field strength component surviving are if $n_1=n_2=1$, giving
$\bar{F}_+$, or if $n_1=-n_2=1$, giving $F_+$.

For the fermions we see that we have one or more fermions if and
only if $n_1+n_3+n_4+|n_5-n_2| = 3$. In particular, we get the
fermion $\chi_1$ if $n_1 - n_2 + n_3 + n_4 + n_5 = 3$ and the
fermion $\bchi_7$ if $n_1+n_2+n_3+n_4-n_5=3$. Having more than one
fermion is only possible in the following cases
\begin{equation}
\label{moreferm}
\begin{array}{lcl}
n = (a,a,1,1,1) &:& \chi_1,\chi_2 \\
n = (1,a,1,1,a) &:& \chi_1,\bchi_7 \\
n = (1,1,1,a,a) &:& \bchi_5,\bchi_7 \\
n = (0,-1,1,1,0)&:& \chi_1, \bchi_8 \\
n = (1,-1,1,0,0) &:& \chi_1,\chi_3 \\
n = (1,1,1,1,1) &:& \chi_1,\chi_2,\bchi_3,\bchi_5,\bchi_7  \\
\end{array}
\end{equation}
where $0 \leq a < 1$. Here we recorded which fermions are present in
each case. We see that we can have either zero, one, two or five
fermions surviving a decoupling limit.

We can now explore systematically what possible number of scalars,
derivatives and fermions can be present in a decoupled theory after
a decoupling limit. Note that there is precisely one gauge field
strength component present if and only if we have two derivatives
present. We begin by considering having 3 scalars and 2 derivatives
present, $i.e.$   the maximally possible number of scalars and
derivatives. In this case, the only limit obeying \eqref{nfercon} is
$n=(1,1,1,1,1)$, and we see from \eqref{moreferm} that this limit
has five fermions present. Consider instead the case of having the
number of scalars plus derivatives equal to four. Taking into
account all the possibilities, it is easily seen that none of them
can obey the constraint \eqref{nfercon}. If the number of scalars
plus derivatives is equal to three it is not hard to see from the
constraints \eqref{nrange} and \eqref{nfercon} that the $n_i$ can
take five different forms, all of them listed in \eqref{moreferm}.
Thus, all of these five possibilities lead to having precisely two
fermions present. Finally, if the number of scalars plus derivatives
is less than or equal to two all possibilities are realized, as one
can see explicitly by our list of decoupling limits below in Section
\ref{sec:declist}. This is with the obvious exception of having zero
scalars, fermions and derivatives, and having one derivative without
any scalar or fermion. Altogether, we obtain 14 different decoupling
limits, with the field content listed in Table \ref{tab:limits}. We
write explicit choices of the $n_i$ for each of the 14 limits below
in Section \ref{sec:declist}.

\begin{table}[ht]
\begin{center}
\begin{tabular}{|l||c|c|c|c|c|c|c|c|c|c|c|c|}
\hline \# derivatives & 0 & 0 & 0 & 0 & 1 & 1 & 1 & 1 & 2 & 2 & 2 & 2 \\
\hline \# scalars     & 0 & 1 & 2 & 3 & 0 & 1 & 2 & 3 & 0 & 1 & 2 & 3 \\
\hline
\hline 0 fermions     &   & + & + &   &   & + &   &   & + &   &   &   \\
\hline 1 fermion     & + & + & + &   & + & + &   &   & + &   &   &   \\
\hline 2 fermions     &   &   &   & + &   &   & + &   &   & + &   &   \\
\hline 5 fermions     &   &   &   &   &   &   &   &   &   &   &   & + \\
\hline
\end{tabular}
\caption{The fourteen possible decoupled theories.
\label{tab:limits}}
\end{center}
\end{table}

\subsection{List of decoupling limits}
\label{sec:declist}

We list here the fourteen possible decoupling limits of $\CN=4$ SYM
on $\R \times S^3$ with gauge group $SU(N)$. The decoupling limits
are all of the form \eqref{declim} and they are specified by the
numbers $(n_1,n_2,n_3,n_4,n_5)$. The fourteen limits give rise to
fourteen different decoupled theories. For each decoupled theory we
give the letter content and we state in which representation of the
symmetry algebra the letters transform. Note that the Dynkin labels
of the algebras used in the following are explained in Appendix
\ref{app:algdecsec}.%
\footnote{See Appendix~\ref{app:oscrep} and Appendix
\ref{app:algdecsec} for more details on the algebras and
representations used in this section.}

\begin{description}
\item[The bosonic $U(1)$ limit.] Given by $n=(0,0,1,0,0)$.
Letter content: $Z$. This limit has previously been considered in \cite{Harmark:2006di}.
\item[The fermionic $U(1)$ limit.]
Given by $n=(\frac{3}{5},-\frac{3}{5},\frac{3}{5},\frac{3}{5},\frac{3}{5})$.
Letter content: $\chi_1$.
\item[The $SU(2)$ limit.] Given by $n=(0,0,1,1,0)$. Letter content: $Z$, $X$.
The letters transform in the $[1]$ representation ($i.e.$  spin
$1/2$ representation) of $su(2)$. This limit has previously been
considered in
Refs.~\cite{Harmark:2006di,Harmark:2006ta,Harmark:2006ie}.
\item[The $SU(1|1)$ limit.] Given by $n=(\frac{2}{3},0,1,\frac{2}{3},\frac{2}{3})$.
Letter content: $Z$, $\chi_1$. The letters transform in the $[1]$ representation of $su(1|1)$.
\item[The $SU(1|2)$ limit.] Given by $n=(\frac{1}{2},0,1,1,\frac{1}{2})$.
Letter content: $Z$, $X$ and $\chi_1$. The letters transform in the $[1,0]$ representation of $su(1|2)$.
\item[The $SU(2|3)$ limit.] Given by $n=(0,0,1,1,1)$. Letter content: $Z$, $X$, $W$, $\chi_1$ and $\chi_2$.
The letters transform in the $[0,0,0,1]$ representation of $su(2|3)$.
This limit has previously been considered in Refs.~\cite{Harmark:2006di,Harmark:2006ta}.
\item[The bosonic $SU(1,1)$ limit.] Given by $n=(1,0,1,0,0)$. Letter content: $d_1^n Z$.
The letters transform in the $[-1]$ representation ($i.e.$  spin
$-1/2$ representation) of $su(1,1)$.
\item[The fermionic $SU(1,1)$ limit.] Given by $n=(1,0,\frac{2}{3},\frac{2}{3},\frac{2}{3})$.
Letter content: $d_1^n \chi_1$. The letters transform in the $[-2]$
representation ($i.e.$  spin $-1$ representation) of $su(1,1)$.
\item[The $SU(1,1|1)$ limit.] Given by $n=(1,0,1,\frac{1}{2},\frac{1}{2})$.
Letter content: $d_1^n Z$ and $d_1^n \chi_1$. The letters transform in the $[0,1]$
representation of $su(1,1|1)$.
\item[The $SU(1,1|2)$ limit.] Given by $n=(1,0,1,1,0)$. Letter content: $d_1^n Z$, $d_1^n X$,
$d_1^n \chi_1$ and $d_1^n \bchi_7$. The letters transform in the $[0,1,0]$ representation of
$su(1,1|2)$.
\item[The $SU(1,2)$ limit.] Given by $n=(1,1,0,0,0)$.
Letter content: $d_1^n d_2^k \bar{F}_+$. The letters transform in
the $[0,-3]$ representation of $su(1,2)$.
\item[The $SU(1,2|1)$ limit.] Given by $n=(1,1,\frac{1}{2},\frac{1}{2},0)$. Letter content:
$d_1^n d_2^k \bar{F}_+$, $d_1^n d_2^k \bchi_7$. The letters transform in the $[0,0,2]$
representation of $su(1,2|1)$.
\item[The $SU(1,2|2)$ limit.] Given by $n=(1,1,1,0,0)$. Letter content:
$d_1^n d_2^k \bar{F}_+$, $d_1^n d_2^k Z$, $d_1^n d_2^k \bchi_5$, $d_1^n d_2^k \bchi_7$.
The letters transform in the $[0,0,0,1]$ representation of $su(1,2|2)$.
\item[The $SU(1,2|3)$ limit.] Given by $n=(1,1,1,1,1)$. Letter content:
$d_1^n d_2^k \bar{F}_+$, $d_1^n d_2^k Z$, $d_1^n d_2^k X$, $d_1^n d_2^k W$,
$d_1^n d_2^k \chi_1$ and $d_1^n d_2^k \chi_2$, $d_1^n d_2^k \bchi_3$,
$d_1^n d_2^k \bchi_5$, $d_1^n d_2^k \bchi_7$.
The letters transform in the $[0,0,0,1,0]$ representation of $su(1,2|3)$.
\end{description}

As explained in Section \ref{sec:systexpl}, the above fourteen
limits constitute a complete list of decoupling limits of the form
\eqref{declim}. There are other possible choices of
$(n_1,n_2,n_3,n_4,n_5)$ that give decoupling limits but the
resulting theories are all equivalent to one of the theories listed
above. For example, the limit given by
$n=(\frac{1}{2},\frac{1}{2},1,1,0)$ gives a decoupled theory
containing $Z, X$ and $\bchi_7$ but this theory is in fact
equivalent to the $SU(1|2)$ theory described above. A few of the
decoupled theories can even be obtained from a continuous family of
choices for $n$. The fermionic $U(1)$ theory can for instance be
found from $n=(a,-a,b,b,b)$ with $0<a,b<1$ satisfying $2a+3b=3$.

The above list of decoupling limits can be divided into the two
trivial limits, being the bosonic and fermionic $U(1)$ limits, and
the twelve non-trivial limits. The twelve non-trivial decoupled
theories can be divided into groups according to the effective
dimensionality of the decoupled theory. The $SU(2)$, $SU(1|1)$,
$SU(1|2)$ and $SU(2|3)$ theories are effectively zero-dimensional so
they correspond to Quantum Mechanical theories. Two of these were
found in \cite{Harmark:2006di}. The bosonic $SU(1,1)$, fermionic
$SU(1,1)$, $SU(1,1|1)$ and $SU(1,1|2)$ theories all have one
derivative present, thus they are effectively one-dimensional.
Finally, the $SU(1,2)$, $SU(1,2|1)$, $SU(1,2|2)$ and $SU(1,2|3)$
theories are effectively two-dimensional, since they each have two
derivatives present.

It is important to note that the above list of limits and theories
are in good correspondence with the list of consistent subgroups of
the $PSU(2,2|4)$ symmetry of $\CN=4$ SYM at the one-loop order, as
examined in \cite{Beisert:2003jj,Beisert:2004ry}. The only exception
is the so-called excitation sector for which the number of
excitations is kept fixed, thus it is not in accordance with our
decoupling limit \eqref{declim}.

\subsection{Closure of $D_2$ in the decoupling limits}
\label{sec:closeD2}

We found above that in the decoupling limit \eqref{declim} for a
given $n=(n_1,n_2,n_3,n_4,n_5)$ only the states with $D_0=J$ survive
and the effective Hamiltonian for the theory becomes $D_0 +
\tilde{\lambda} D_2$. In the following we show that this is
consistent with the $D_2$ operator.

We begin by reviewing briefly the $D_2$ operator, as found by
Beisert \cite{Beisert:2003jj}. The $D_2$ operator acts on two
letters at a time in a given operator. We can therefore think of
$D_2$ in terms of the action on $\CA \times \CA$, $i.e.$ on the
product of two singleton representations of $psu(2,2|4)$. It is
found that $\CA \times \CA$ splits up in a sum of representations as
follows
\begin{equation}
\CA \times \CA = \sum_{j=0}^\infty \CV_j
\end{equation}
where the singleton representation $\CA$ and the modules $\CV_j$ are
\cite{Dolan:2002zh,Bianchi:2003wx}
\begin{equation}
\mathcal{A}=\mathcal{B}^{\frac{1}{2},\frac{1}{2}}_{[0,1,0]_{(0,0)}}\quad
\CV_0=\mathcal{B}^{\frac{1}{2},\frac{1}{2}}_{[0,2,0]_{(0,0)}}\quad
\CV_1=\mathcal{B}^{\frac{1}{4},\frac{1}{4}}_{[1,0,1]_{(0,0)}}\quad
\CV_j=\mathcal{C}^{1,1}_{[0,0,0]_{(\frac{j}{2}-1,\frac{j}{2}-1)}}~\mbox{for
$j\geq 2$} \label{modules}
\end{equation}
written in the notation of \cite{Dolan:2002zh}, where for each
module it is specified which superconformal primary operator the
representation is generated from. With this we can write the $D_2$
operator as \cite{Beisert:2003jj}
\begin{equation}
\label{d2op} D_2 = - \frac{1}{2 N} \sum_{j=0}^\infty h(j)
(P_j)^{AB}_{CD} : \mbox{Tr} [ W_A , \bar{W}^C ] [W_B , \bar{W}^D] :
\end{equation}
where $h(j) = \sum_{k=1}^j \frac{1}{k}$
are the harmonic numbers, $P_j$ is the projection operator
to the module $\CV_j$ and $W_A$ represent all possible letters of
$\CN=4$ SYM.

The $D_2$ operator \eqref{d2op} commutes by construction with all
the generators of the tree-level superconformal algebra $psu(2,2|4)$
(see Appendix \ref{app:oscrep}) \cite{Beisert:2003jj}. In
particular, this means that
\begin{equation}
\label{D2coms} [D_2,D_0] = 0, \quad
[D_2,S_a] = 0, \quad
[D_2,J_i] = 0
\end{equation}
with $a=1,2$ and $i = 1,2,3$. As a consequence of this, we see that
\begin{equation}
\label{d2J} [D_2, D_0 - J] = 0
\end{equation}
with $J$ as defined in \eqref{defJ}.

Using Eq.~\eqref{d2J} we can now show that $D_2$ is closed in any of
the decoupled theories listed in Section \ref{sec:declist}. For a
decoupling limit with a given $n=(n_1,n_2,n_3,n_4,n_5)$ the states
in the corresponding decoupled theory are the ones with $D_0 - J =
0$. Therefore, Eq.~\eqref{d2J} means that the decoupled theory is
closed with respect to $D_2$ since the action of $D_2$ on any state
with $D_0-J=0$ will give a new state with $D_0-J=0$.

\section{Spectrum of decoupled theories in planar limit}
\label{sec:planar}

In this section we consider the decoupled theories found in Section
\ref{sec:newlim} in the planar limit. In the planar limit it is
possible to single out the single-trace operators, and the spectrum
of the multi-trace operators can be found from the knowledge of the
spectrum of the single-trace operators. Using furthermore the spin
chain interpretation for the single-trace operators
\cite{Minahan:2002ve} it is possible to find a Bethe equation that
contains the full spectrum of the effective Hamiltonian $D_0 +
\tilde{\lambda} D_2$. We review how this works, and we use this to
obtain explicitly the low energy spectrum for the decoupled theories
found in Section \ref{sec:newlim} in the planar limit.

Note that the technology used to find the spectrum of $D_2$ has been
developed mainly in \cite{Minahan:2002ve,Beisert:2003tq,Beisert:2003jj,Beisert:2003yb}. In this section we apply this technology to derive the specific spectra for the effective Hamiltonian $D_0 + \tilde{\lambda} D_2$.

\subsection{Full spectrum from Bethe equations}

In the planar limit of $\CN=4$ SYM, a single-trace operator with $L$ letters
\begin{align}
\tr\left(A_1A_2\cdots A_L\right)
\end{align}
can be interpreted as a state of a periodic homogenous spin chain of
length $L$ where each letter in the trace corresponds to a spin in
one site of the spin chain \cite{Minahan:2002ve}.  The simplest
example of this correspondence is the $SU(2)$ sector which contains
only two types of letters, $Z$ and $X$, corresponding to the spin-up
and spin-down states in the spin chain. The dynamics of the spin
chain is governed by a Hamiltonian which in our case is
$D_0+\tilde\lambda D_2$. The spectrum of $D_2$ for $\CN=4$ SYM on
$\R\times S^3$ in the planar limit is given by the $PSU(2,2|4)$
super spin chain found in \cite{Beisert:2003yb}. In the following we
use this to find the spectrum of $D_2$ in the planar limit for the
various decoupled theories.

For the decoupled theories that contain one or more of
the complex scalars $Z$, $X$ and $W$,  the vacuum sector
consists of the symmetrized combinations of the scalars, e.g.\ for
the $SU(2)$ theory the vacuum states are of the form $\tr (
\mbox{sym} (Z^m X^n) )$. The value of $D_2$ on such states is zero,
which is connected to the fact that these particular single-trace
operators correspond to chiral primaries in $\CN=4$ SYM. See
\cite{Harmark:2006ta} for a discussion of this for the $SU(2)$
theory.

There are three decoupled theories which do not contain any of the scalars $Z$,
$X$ and $W$, and for these the $D_2$ vacuum energy is
shifted from zero. The fermionic $SU(1,1)$ theory has ground state
$\tr (\chi_1^L)$ with $D_2$ eigenvalue $L$, the $SU(1,2)$ theory has
ground state $\tr (\bar F_+^L)$ with $D_2$ eigenvalue $3L/2$ and the
$SU(1,2|1)$ theory has ground state $\tr(\bar\chi_7^L)$ with $D_2$
eigenvalue $L$. As we explain below and in Section \ref{sec:largelambsu12},
this has important implications for considering the large
$\tilde{\lambda}$ limit.

The effective Hamiltonian for our decoupled theories is
\begin{align}
\label{eq:effHamiltonian}
H = D_0 + \tilde\lambda D_2.
\end{align}
The Bethe ansatz technique is only relevant for the $D_2$ part of
the Hamiltonian. Instead, for the $D_0$ part we use that any
eigenstate of the spin chain is an eigenstate of $D_0$. In general
the $D_0$ eigenvalue will depend on the excitations of the spin
chain. This dependence can in many cases be interpreted as a Zeeman
coupling to an external magnetic field, as we shall see below in
Section \ref{sec:lowspec}.

The spin chains that correspond to the planar limit of the decoupled
theories found in Section~\ref{sec:newlim} are all integrable and the
spectrum of $D_2$ for each of them is determined by using the Bethe
ansatz technique \cite{Beisert:2004ry}. By using the Dynkin labels
$V_a$ and Cartan matrix $M_{ab}$ of each decoupled theory (see
Appendix \ref{app:algdecsec}), we can
treat them at the same time and obtain the spectrum of $D_2$ from
the generalized Bethe equation.
Each eigenstate of $D_2$ is determined by a set of Bethe roots
$u_k$, $k=1,...,K$, where $K$ is the total number of excitations.
Some of our decoupled theories have a symmetry algebra of rank
higher than one and for these theories it is important to specify
which simple root of the Dynkin diagram each Bethe excitation
corresponds to. This is done with the label $j_k$ which for each
Bethe excitation can take values from one and up to the rank of the
symmetry algebra.

The eigenvalue of $D_2$ on a state with $K$ excitations is given by
\cite{Beisert:2003yb}
\begin{align}
\label{eq:genBetheSpectrum}
D_2= \frac{1}{2} \sum_{k=1}^K \frac{|V_{j_k}|}{u_k^2 + \frac{1}{4}V_{j_k}^2} + c L
\end{align}
where we have included the possible shift $cL$ with $c\in\{0,1,3/2\}$,
depending on the ground state of the theory as discussed above.
It turns out that our decoupled theories all have the property that only one of the
Dynkin labels is non-vanishing.  Therefore only excitations corresponding
to this one non-zero Dynkin label will give contributions to the spectrum.

The Bethe roots are determined by the general Bethe equations that can
be written in compact form as \cite{Beisert:2003yb,Beisert:2004ry}
\begin{align}
\label{eq:genBethe1}
\left(\frac{u_k+ \frac{i}{2}V_{j_k}}{u_k- \frac{i}{2}V_{j_k}} \right)^L
= \prod_{\ell=1,\ell \ne k}^K
\left(\frac{u_k - u_\ell + \frac{i}{2}M_{j_k,j_\ell}}
{u_k - u_\ell - \frac{i}{2}M_{j_k,j_\ell}} \right)
\end{align}
with the cyclicity condition
\begin{align}
\label{eq:genBethe2}
U = \prod_{k=1}^K
\left(\frac{u_k + \frac{i}{2}V_{j_k}}{u_k - \frac{i}{2}V_{j_k}} \right)
\end{align}
where $U=1$ for the decoupled theories with bosonic vacua and
$U=(-1)^L$ for the two decoupled theories in which we have a
fermionic vacuum state.
The full spectrum of the effective Hamiltonian \eqref{eq:effHamiltonian}
in the planar limit is determined by
Eqs.~\eqref{eq:genBetheSpectrum}--\eqref{eq:genBethe2}
for all decoupled theories.

Some of the decoupled theories considered here are well known in the
Condensed Matter literature.  The $SU(2)$ theory is for example
equivalent to the Heisenberg $XXX_{1/2}$ model while the bosonic
$SU(1,1)$ theory is the non-compact $XXX_{-1/2}$ Heisenberg model
and the fermionic $SU(1,1)$ is the non-compact spin $-1$ $XXX$ model
\cite{Beisert:2004ry}.
The $SU(1|1)$ theory is equivalent to a Heisenberg $XX_{1/2}$ spin
chain in an external magnetic field which describes free fermions
and is exactly solvable. We will discuss this theory further in
Section~\ref{sec:SU11asXX}.
Finally, the $SU(1|2)$ theory is equivalent to the so called $t-J$
model \cite{Beisert:2005fw} that is believed to be relevant for high
$T_c$ superconductivity.

We see thus, that our decoupling limits \eqref{declim} for planar
$\CN=4$ SYM on $\R \times S^3$ lead to known Condensed Matter
theories which are fully decoupled. In other words, when approaching
certain of the critical points found in Section \ref{sec:newlim},
planar $\CN=4$ SYM on $\R\times S^3$ reduces to known Condensed
Matter theories.


\subsection{Low energy spectrum in the thermodynamic limit}
\label{sec:lowspec}

It is in general hard to solve explicitly the Bethe equations
\eqref{eq:genBethe1}--\eqref{eq:genBethe1}, but we
can easily obtain a leading order solution for the low energy spectrum in the
thermodynamic limit $L\to\infty$.  In this regime the positions
of the roots $u_k$ scale like $L$ \cite{Minahan:2002ve} and we therefore
define $u_k = L \tilde u_k$.
Plugging this into Eq.~\eqref{eq:genBethe1} and taking the logarithm,
we find
\begin{align}
\label{eq:genBetheL}
2\pi n_k -  \frac{V_{j_k}}{\tilde u_k}
= \frac{1}{L}\sum_{\ell=1,\ell\ne k}^{K} \frac{M_{j_k,j_\ell}}{\tilde u_\ell - \tilde u_k} + \CO(L^{-2}),
\end{align}
where $n_k$ are integers. Neglecting the right hand side to
leading order in $1/L$, Eq.~\eqref{eq:genBetheL} gives the solution
\begin{align}
\label{eq:leadingsolution}
\tilde u_k = \frac{V_{j_k}}{2\pi n_k} + \CO(L^{-1})
\end{align}
and inserting that into the spectrum \eqref{eq:genBetheSpectrum} we
obtain
\begin{align}
\label{eq:genlowspectrum}
D_2 = \frac{2\pi^2}{L^2} \sum_{k=1}^{K'} \frac{n_k^2}{|V_{j_k}|}
  +  \CO(L^{-3})
\end{align}
where the sum now only goes over the Bethe roots that correspond
to the simple root of the Dynkin diagram with non-vanishing Dynkin
label.

Plugging the leading order solution \eqref{eq:leadingsolution}
into the constraint equation \eqref{eq:genBethe2} gives
\begin{align}
\sum_{k=1}^{K'} n_k = 0
\end{align}
which is the zero-momentum condition for the spin chain and the cyclicity
condition for the trace on the gauge theory side.
For bosonic excitations we can have more than one excitation with the
same $n_k$, whereas for fermionic excitations we can at most have
one excitation with a given value of $n_k$.
We must therefore distinguish between scalar excitations, derivatives and
fermionic excitations.
For the two possible scalar excitations, we denote the number of $n_k$
that are equal to a particular integer $n$ as $M_n^{(i)}$, $i=1,2$,
for the two derivative excitations we denote the number as
$N_n^{(j)}$, $j=1,2$, and for the four possible fermionic excitations as
$F_n^{(\alpha)}$, $\alpha =1,...,4$.
From the oscillator representation in Appendix~\ref{app:algdecsec}
we can see that not all excitations are independent,
$\bar F_+$ is for example a composite field and we do not need to
keep track of it in the partition function.  The same is true for $\bchi_3$ which is
composed of the $\bchi_7$ and $W$ excitations.
Therefore we only have four different types of fermionic excitations and
not five as one might guess in a theory with five fermions like the $SU(1,2|3)$.

\subsubsection*{All decoupled theories containing scalars}

Nine out of the 12 non-trivial decoupled theories found in Section~\ref{sec:newlim}
contain at least one scalar and their spectra can all be described in the same
way using the number operators $M_n$, $N_n$ and $F_n$.
Depending on their letter content, the decoupled theories have different number of
these operators appearing and in Table~\ref{tab:abc} we list
how many there are of each of the three possible types.
\begin{table}
\begin{center}
\begin{tabular}{c|ccccccccc}
$SU(\cdot)$&$(2)$ & $(1,1)_\textrm{bos}$ & $(1|1)$ & $(1|2)$ & $(2|3)$ & $(1,1|1)$ & $(1,1|2)$ & $(1,2|2)$ & $(1,2|3)$ \\
\hline
$a$&1&0&0&1&2&0&1&0&2\\
$b$&0&1&0&0&0&1&1&2&2\\
$c$&0&0&1&1&2&1&2&2&4
\end{tabular}
\caption{The table shows how many number operators we have of each
type ($a$ for scalars $M_n$, $b$ for derivatives $N_n$, and $c$ for
fermions $F_n$) in each of the nine theories that contain at least
one scalar. $SU(1,1)_\textrm{bos}$ corresponds to the bosonic $SU(1,1)$ theory.
The numbers in this table will be used again in Section~\ref{sec:largelambda}
where we consider the partition functions for large $\tilde\lambda$.
\label{tab:abc}}
\end{center}
\end{table}
These theories all share the feature that the absolute value of the
single non-zero Dynkin label is equal to one and therefore the
spectra for these nine different theories all take the form
\begin{align}
\label{eq:ABCspectrum}
H  = L
+ \sum_{n\in \mathbb{Z}} \left(
\sum_{j=1}^{b} N_n^{(j)}
+ \frac{1}{2}  \sum_{\alpha=1}^c F^{(\alpha)}_n
 \right)
 + \frac{2\pi^2\tilde\lambda}{L^2} \sum_{n\in \mathbb{Z}} n^2
\left(
\sum_{i=1}^a M_n^{(i)} +\sum_{j=1}^b N_n^{(j)} + \sum_{\alpha=1}^c F_n^{(\alpha)}
\right)
\end{align}
with the cyclicity (zero momentum) constraint
\begin{align}
\label{eq:ABCconstraint}
P \equiv \sum_{n\in \mathbb{Z}} n \left(
\sum_{i=1}^a M_n^{(i)} +\sum_{j=1}^b N_n^{(j)}
+ \sum_{\alpha=1}^c F_n^{(\alpha)} \right) = 0.
\end{align}
Note that $F_n^{(\alpha)} \in \{0,1\}$ while
$M_n^{(i)}, N_n^{(j)} \in \{0,1,2,...\}$. The numbers $a,b$ and $c$
are given in Table~\ref{tab:abc}.

The first two terms in the spectrum come from $D_0$. Recall that
in the decoupling limit we have that $D_0 = J$.
The vacuum is made from the scalars which all contribute $1$ to $J$
and therefore the vacuum has $J=L$.
Each derivative gives an additional contribution and
$\sum_{n\in\mathbb{Z}}\sum_{j=1}^{b} N_n^{(j)}$ precisely counts the
number of derivatives.  Similarly
$\sum_{n\in\mathbb{Z}}\sum_{\alpha=1}^c F^{(\alpha)}_n$
counts the total number of fermions and each of them contributes
$1/2$ more to $J$ than the scalars do.
The second term can be interpreted as a coupling of the spin chain
to an external magnetic field through a Zeeman term \cite{Harmark:2006ie}.

\subsubsection*{Decoupled theories without scalars}

The three decoupled theories without scalars have their $D_2$
vacuum shifted from zero.  We can still use the Bethe ansatz to find their
low energy spectrum, but it will not be as useful to us when we consider
the large $\tilde\lambda$ Hagedorn temperature.

The fermionic $SU(1,1)$ theory is the simplest example of a decoupled
theory with a non-vanishing $D_2$ vacuum energy.
The ground state is made from fermions and we assume that $L$ is odd to satisfy
the cyclicity constraint.  Since $\chi_1$ is now the
highest weight, the representation has Dynkin label $V=-2$ and this
theory is equivalent to the Heisenberg $XXX_{-1}$ spin chain \cite{Beisert:2004ry}.
The spectrum of the Hamiltonian is
\begin{align}
\label{eq:Hsu11f}
H = \left(\frac{3}{2} + \tilde\lambda\right) L
+\sum_{n\in\mathbb{Z}} N_n^{(1)}
+\frac{\pi\tilde\lambda}{L^2} \sum_{n\in\mathbb{Z}} n^2 N_n^{(1)}
\end{align}
where we note that $\tilde\lambda$ already appears in the first term.
We also have the usual zero-momentum constraint analogous to
Eq.~\eqref{eq:ABCconstraint}.

The $SU(1,2)$ theory is very interesting since it shares many features
with QCD, as discussed in Section \ref{sec:pureYM}.  The highest weight is $\bar F_+$ and the representation has
Dynkin label $V=[0,-3]$.  The spectrum of the Hamiltonian is
\begin{align}
\label{eq:Hsu12}
H = \left(2 + \frac{3}{2}\tilde\lambda\right) L
+\sum_{n\in\mathbb{Z}} \left(N_n^{(1)} + N_n^{(2)} \right)
+\frac{2\pi\tilde\lambda}{3L^2}
   \sum_{n\in\mathbb{Z}} n^2 \left(N_n^{(1)}+N_n^{(2)}\right).
\end{align}

The third theory that does not contain any scalar is $SU(1,2|1)$ where the
highest weight is again $\chi_1$ and the spectrum can straightforwardly
be worked out along similar lines as for the fermionic $SU(1,1)$ and the
$SU(1,2)$ decoupled theories.

The $\tilde\lambda L$ term that appears in the spectrum of these decoupled
theories has important implications when $\tilde\lambda$ and $L$ are large.
This will be discussed in Sections~\ref{sec:largelambsu12} and \ref{sec:stringreg}.


\section{Finite temperature behavior in planar limit}
\label{sec:temp}

In this section we begin by generalizing the computation of the partition
function for free $\CN=4$ SYM on $\R \times S^3$
\cite{Sundborg:1999ue,Polyakov:2001af,Aharony:2003sx,Yamada:2006rx,Harmark:2006di} to include
all five possible chemical potentials. Applying then the decoupling limit
\eqref{declim} we find the partition function and Hagedorn temperature for each of the decoupled theories at zero coupling.

Turning on a small 't Hooft coupling $\lambda$, we compute in
Section \ref{sec:smalllambda} the one-loop correction to the
Hagedorn temperature for small values of $\tilde\lambda$ in all the
decoupled theories. The procedure is a generalization of the one
used in~\cite{Spradlin:2004pp,Harmark:2006di}.

In Section \ref{sec:largelambda} we compute the Hagedorn temperature in the
large $\tilde\lambda$ regime. This is done by using a general relation
between the Hagedorn temperature in the decoupled theories and the
free energy per site of the corresponding spin chain model in the
thermodynamic limit \cite{Harmark:2006ta}. We use this method for
all the decoupled theories containing scalars.

In section \ref{sec:SU11asXX} we examine the $SU(1|1)$ theory. We
compute the one-loop Hagedorn temperature for all values of
$\tilde\lambda$ using the relation with the Heisenberg $XX_{1/2}$
spin chain.

Finally in section \ref{sec:largelambsu12} we study the large
$\tilde\lambda$ regime of the $SU(1,2)$ theory which is
particularly interesting for its connection to pure Yang-Mills theory.


\subsection{Partition function of the free theory}
\label{sec:freetheory}

In this section we compute of the partition
function for free $\CN=4$ $SU(N)$ SYM on $\R \times S^3$ with
all five possible chemical potentials turned on.
The partition function of free $SU(N)$ SYM on $\R \times S^3$ can be found from the letter partition function \cite{Sundborg:1999ue,Polyakov:2001af,Aharony:2003sx}. With chemical potentials turned on, the only difference is that one needs the letter partition function with chemical potentials \cite{Aharony:2003sx,Yamada:2006rx,Harmark:2006di}. Below we compute the general letter partition function $z(x,\rho_j,y_i)$ depending on the temperature and all five chemical potentials, where we introduce the notation
\begin{equation}
x \equiv e^{-\beta} \spa \rho_j \equiv e^{\beta \omega_j} \,, j=1,2 \spa y_i \equiv e^{\beta \Omega_i} \,, i=1,2,3 \ .
\end{equation}
With the letter partition function $z(x,\rho_j,y_i)$ one can then find the full partition function for free $SU(N)$ $\CN=4$ SYM on $\R \times S^3$ as
\begin{equation}
\label{fullpart}
Z_{\lambda=0,N}(x,\rho_j,y_i) = \int [dU] \exp
\left[ \sum_{k=1}^\infty \frac{1}{k} z(\eta^{k+1} x^k,\rho_j^k,y_i^k) \left( \tr (U^k) \tr (U^{-k}) - 1 \right) \right]
\end{equation}
where $\eta=e^{2\pi i}$ is introduced to take the correct sign into
account for the fermions. In the planar limit $N=\infty$, the partition function
\eqref{fullpart} becomes
\begin{equation}
\label{multinfty} \log Z_{\lambda=0,N=\infty}(x,\rho_j,y_i) = - \sum_{k=1}^\infty \log
\left[ 1 - z(\eta^{k+1} x^k,\rho_j^k,y_i^k) \right] \,.
\end{equation}

One can see from \eqref{multinfty} that one encounters a singularity when $z(x,\rho_j,y_i)=1$. This is the Hagedorn singularity for planar $\CN=4$ $SU(N)$ SYM on $\R \times S^3$ \cite{Sundborg:1999ue,Polyakov:2001af,Aharony:2003sx,Yamada:2006rx,Harmark:2006di}. With chemical potentials, we see that the equation $z(x,\rho_j,y_i)=1$ defines the Hagedorn temperature $T_{\rm H} (\omega_j,\Omega_i)$ as a function of all the five chemical potentials.%
\footnote{Note that this means that \eqref{multinfty} is only valid
for temperatures below the Hagedorn temperature $T_{\rm H}
(\omega_j,\Omega_i)$. If we want to study the theory above the
Hagedorn temperature we need to go beyond the planar limit. This is 
in accordance with the fact that the Hagedorn temperature $T_{\rm H}
(\omega_j,\Omega_i)$ is limiting for free $\CN=4$ SYM on $\R \times
S^3$ in the planar limit $N=\infty$, thus it takes an infinite
amount of energy to reach the Hagedorn temperature.}

In the following we first compute the full letter partition function
of $\CN=4$ SYM on $\R \times S^3$ with all five chemical potentials
turned on. Then we take the decoupling limit \eqref{declim} of the
obtained letter partition function of $\CN=4$ SYM, thus finding the
partition function for each of the decoupled theories in the free
limit.\footnote{See also \cite{Bianchi:2006ti,Dolan:2007rq}.} 
Employing the letter partition function for the decoupled
theories we furthermore compute the Hagedorn temperature for each
decoupled sector in the free limit. Note that the results of the
computations for the decoupled theories are listed at the end of
section \ref{sec:smalllambda}.

\subsubsection*{Computing the letter partition function}

We compute now the letter partition function
for $\CN=4$ SYM on $\R \times S^3$ in the presence of non-zero
chemical potentials for the R-charges of the $SU(4)$ R-symmetry and
for the Cartan generators of the $SO(4)$ symmetry group of $S^3$.
To compute the letter partition function we use the spherical
harmonic expansion method by expanding each field in the spectrum of $\CN=4$ SYM
in terms of the corresponding spherical harmonics. To do this,
instead of  the Cartan generators of the $SO(4)$ symmetry $S_1$ and
$S_2$, it is convenient to define the operators
\begin{equation}
S_L=\frac{S_1-S_2}{2},\quad S_R=\frac{S_1+S_2}{2} \label{slsr}
\end{equation}
corresponding to the generators of the $SU(2)_L\times SU(2)_R$
symmetry.


We begin with the scalars.
The spherical harmonics corresponding to scalars are denoted by
$S_{j,m,\bar m}(\alpha)$, where $\alpha$ represents the coordinates
of $S^3$ and $m$, $\bar m$ label the eigenvalues of $S_L$ and $S_R$
respectively. Their values are $m=\bar m=-j/2,
-j/2+1,...,j/2-1,j/2$.

From Table~\ref{tab:scalars}, we see that all the six scalars are
in the same representation of $SU(2)_L\times SU(2)_R$. The scalar
partition function can therefore be written as
\begin{equation}
\eta_S(x,\rho,\bar\rho,y_i)=
\sum_{i=1}^3\sum_{j=0}^{\infty}\sum_{m=-j/2}^{j/2}
\sum_{\bar m=-j/2}^{j/2}x^{j+1}
\rho^m\bar \rho^{\bar m}(y_i+y_i^{-1})\label{sca}
\end{equation}
where we introduced the notation
\begin{equation}
\label{rhobar} \rho \equiv e^{\beta (\omega_1-\omega_2)},\quad \bar
\rho \equiv e^{\beta (\omega_1+\omega_2)}.
\end{equation}
Performing the sums, we get the following result for the scalar
partition function
\begin{equation}
\eta_S(x,\omega_j,y_i)=\frac{(x-x^3)}{(1-xe^{\beta\omega_1})
(1-xe^{-\beta\omega_1}) (1-xe^{\beta\omega_2})
(1-xe^{-\beta\omega_2})}\sum_{i=1}^3\left(y_i+y_i^{-1}\right).
\label{scalar}
\end{equation}
%


Turning to the vectors, we have that they are neutral under the R-charges. The spherical harmonics
corresponding to the gauge boson in the representation
$[0,0,0]_{(1,0)}$ are denoted by $V^{L}_{j,m,\bar m}(\alpha)$ with
$m=-(j+1)/2, ...,(j+1)/2$ and $\bar m=-(j-1)/2, ...,(j-1)/2$. Their
contribution to the letter partition function is given by
\begin{equation}
\eta_{V^L}(x,\rho,\bar
\rho,y_i)=\sum_{j=1}^{\infty}\sum_{m=-(j+1)/2}^{(j+1)/2}\sum_{\bar
m=-(j-1)/2}^{(j-1)/2}x^{j+1}\rho^m\bar \rho^{\bar m}\label{vl}.
\end{equation}
The spherical harmonics corresponding to the gauge boson in the
representation $[0,0,0]_{(0,1)}$ are denoted by $V^{R}_{j,m,\bar
m}(\alpha)$ with $m=-(j-1)/2, ...,(j-1)/2$ and $\bar m=-(j+1)/2,
...,(j+1)/2$. Their contribution to the letter partition function is
given by
\begin{equation}
\eta_{V^R}(x,\rho,\bar
\rho,y_i)=\sum_{j=1}^{\infty}\sum_{m=-(j-1)/2}^{(j-1)/2}\sum_{\bar
m=-(j+1)/2}^{(j+1)/2}x^{j+1}\rho^m\bar \rho^{\bar m}\label{vr}.
\end{equation}
Performing the sums and adding together the two contributions, we
get the following result for the vector partition function
\begin{equation}
\eta_V(x,\omega_j)=\frac{2x^2\left[1+2\cosh (\beta \omega_1) \cosh
(\beta \omega_2)-2x \left(\cosh (\beta \omega_1)+\cosh (\beta
\omega_2)\right)+x^2\right]}{(1-xe^{\beta\omega_1})
(1-xe^{-\beta\omega_1}) (1-xe^{\beta\omega_2})
(1-xe^{-\beta\omega_2})} \label{vector}.
\end{equation}
%


Finally, we turn to the fermions.
Fermions appear in two representations, $[0,0,1]_{(1/2,0)}$ and
$[1,0,0]_{(0,1/2)}$. For the representation $[0,0,1]_{(1/2,0)}$ we
can introduce the spherical harmonics $F^{1}_{j,m,\bar m}(\alpha)$
with $m=-(j)/2, ...,(j)/2$ and $\bar m=-(j-1)/2, ...,(j-1)/2$.
Taking into account the dependence on the R-charge chemical
potentials for fermions in this representation which is given by
\begin{equation}
Y_1=(y_1y_2y_3)^{1/2}+y_1^{1/2}(y_2y_3)^{-1/2}+(y_1y_3)^{-1/2}y_2^{1/2}+(y_1y_2)^{-1/2}y_3^{1/2}
\label{y1}
\end{equation}
we obtain the letter partition function
\begin{equation}
\eta_{F^1}(x,\rho,\bar
\rho,y_i)=Y_1\,\sum_{j=1}^{\infty}\sum_{m=-(j)/2}^{(j)/2}\sum_{\bar
m=-(j-1)/2}^{(j-1)/2}x^{j+\frac{1}{2}}\rho^m\bar \rho^{\bar m}.
\end{equation}
The result in terms of $\omega_1$ and $\omega_2$ is given by
\begin{equation}
\eta_{F^1}(x,\omega_j,y_i)=Y_1\,\frac{2x^{3/2}\left(\cosh
\left[\beta \left(\frac{\omega_1-\omega_2}{2}\right)\right]-x \cosh
\left[\beta
\left(\frac{\omega_1+\omega_2}{2}\right)\right]\right)}{(1-xe^{\beta\omega_1})
(1-xe^{-\beta\omega_1}) (1-xe^{\beta\omega_2})
(1-xe^{-\beta\omega_2})} \label{f1}.
\end{equation}

For fermions in the representation $[1,0,0]_{(0,1/2)}$, we can
introduce the spherical harmonics $F^{2}_{j,m,\bar m}(\alpha)$ with
$m=-(j-1)/2, ...,(j-1)/2$ and $\bar m=-(j)/2, ...,(j)/2$. The
dependence on the R-charge chemical potentials in this case is given
by
\begin{equation}
Y_2=(y_1y_2y_3)^{-1/2}+y_1^{-1/2}(y_2y_3)^{1/2}+(y_1y_3)^{1/2}y_2^{-1/2}+(y_1y_2)^{1/2}y_3^{-1/2}
\label{y2}
\end{equation}
and the contribution to the letter partition function is
\begin{equation}
\eta_{F^2}(x,\rho,\bar
\rho,y_i)=Y_2\,\sum_{j=1}^{\infty}\sum_{m=-(j-1)/2}^{(j-1)/2}\sum_{\bar
m=-(j)/2}^{(j)/2}x^{j+\frac{1}{2}}\rho^m\bar \rho^{\bar m}.
\end{equation}
The result in terms of $\omega_1$ and $\omega_2$ is given by
\begin{equation}
\eta_{F^2}(x,\omega_j,y_i)=Y_2\,\frac{2x^{3/2}\left(\cosh
\left[\beta \left(\frac{\omega_1+\omega_2}{2}\right)\right]-x \cosh
\left[\beta
\left(\frac{\omega_1-\omega_2}{2}\right)\right]\right)}{(1-xe^{\beta\omega_1})
(1-xe^{-\beta\omega_1}) (1-xe^{\beta\omega_2})
(1-xe^{-\beta\omega_2})} \label{f2}.
\end{equation}
%


Adding together the contributions of scalars, vectors and fermions,
we obtain the letter partition function for $\CN=4$ SYM on $\R
\times S^3$ in the presence of non-zero chemical potentials for the
R-charges of the $SU(4)$ R-symmetry and for the Cartan generators of
the $SO(4)$ symmetry group of $S^3$ which is given by
\begin{align}
\label{letter} z\left(x,\omega_j,y_i\right) =&\prod_{k=1}^2\left(
 (1-xe^{\beta\omega_k})
(1-xe^{-\beta\omega_k}) \right)^{-1} \bigg\{
(x-x^3)\sum_{l=1}^3(y_l+y_l^{-1}) \nn\\
&+2x^2\left[1+2\cosh (\beta \omega_1) \cosh (\beta \omega_2)
-2x \left(\cosh (\beta \omega_1)+\cosh (\beta
\omega_2)\right)+x^2\right] \nn\\
&+Y_1\,2x^{3/2}\left[\cosh [\beta
(\frac{\omega_1-\omega_2}{2})]-x \cosh [\beta
(\frac{\omega_1+\omega_2}{2})]\right] \nn\\
&+Y_2\,2x^{3/2}\left[\cosh [\beta
(\frac{\omega_1+\omega_2}{2})]-x \cosh [\beta
(\frac{\omega_1-\omega_2}{2})]\right]\bigg\}.
\end{align}
As shown in Appendix~\ref{sec:freetheory2}, the above result for the letter partition function can also be obtained using the oscillator representation of $\CN=4$ SYM
\cite{Gunaydin:1984fk,Beisert:2003jj}.

With the letter partition function \eqref{letter} in hand, the
partition function of free $\CN=4$ SYM on $\R \times S^3$ with all
five possible chemical potentials turned on is given by
Eq.~\eqref{fullpart}, or Eq.~\eqref{multinfty} in the planar limit.

\subsubsection*{Free partition functions for the decoupled theories}

We can now find the partition function for each of the decoupled theories when $\tilde{\lambda}=0$ by taking the decoupling limit \eqref{declim}. This is done by taking the decoupling limit of the letter partition function \eqref{letter}. Defining $Y \equiv \exp (i \beta \Omega)$, we can write the decoupling limit of the letter partition function as
\begin{equation}
x\to 0~,\ Y\to\infty \quad \mbox{with $\tilde x=xY$
fixed}\label{fredeclim} \,.
\end{equation}
Given one of the $n=(n_1,n_2,n_3,n_4,n_5)$ for the fourteen
decoupling limits listed in Section \ref{sec:declist}, we set the
chemical potentials to be given by \eqref{omegas}, and then take the
limit \eqref{fredeclim} of the letter partition function
\eqref{letter}. The resulting letter partition functions for the
twelve non-trivial decoupled theories are listed at the end of
Section \ref{sec:smalllambda}. Given one of the decoupled letter
partition functions $z(\tilde{x})$ we can then find the partition
function for free $SU(N)$ $\CN=4$ SYM on $\R \times S^3$ in the
decoupling limit as
\begin{equation}
\label{fulldecpart}
Z_{\tilde{\lambda}=0,N}(\tilde{x}) = \int [dU] \exp
\left[ \sum_{k=1}^\infty \frac{1}{k} z(\eta^{k+1} \tilde{x}^k) \left( \tr (U^k) \tr (U^{-k}) - 1 \right) \right] \,.
\end{equation}
In the planar limit $N=\infty$ this reduces to
\begin{equation}
\label{decmultinfty} \log Z_{\tilde{\lambda}=0,N=\infty}(\tilde{x}) = - \sum_{k=1}^\infty \log
\left[ 1 - z(\eta^{k+1} \tilde{x}^k) \right] \,.
\end{equation}

We see from \eqref{decmultinfty} that we have a Hagedorn singularity for $z(\tilde{x})=1$. This defines the Hagedorn temperature $\tilde{T}^{(0)}_{\rm H}$ for each of the twelve non-trivial decoupled theories. In the end of Section~\ref{sec:smalllambda} we have listed $\tilde{T}^{(0)}_{\rm H}$ for each of the theories.%
\footnote{
It is interesting to notice that some theories have
the same free Hagedorn temperature and that the chemical potentials
in these theories are all related by a permutation.  The theories
$SU(1|2)$, $SU(1,1|1)$ and $SU(1,2|1)$ have for example all the
same $\tilde T_\textrm{H}^{(0)}$ and their critical chemical potentials
$(n_1,n_2,n_3,n_4,n_5)$ are all given with some permutation of
$(1,1,\frac{1}{2},\frac{1}{2},0)$.} For the two $U(1)$ theories $\tilde T$ can be
arbitrarily large and the Hagedorn singularity is never reached.


\subsection{Hagedorn temperature for small $\tilde\lambda$}
\label{sec:smalllambda}

In this section we consider small $\tilde\lambda$ and work out the
Hagedorn temperature up to one-loop order for each of the decoupled
theories. The results are presented as a list at the end of this section.

The general formula for the one-loop correction to Hagedorn
temperature is given by~\cite{Spradlin:2004pp, Harmark:2006di}
\begin{equation}
\delta \tilde{T}_\textrm{H} = \tilde \lambda \left. \frac{ \langle
D_2\left(\tilde x\right) \rangle }{ \tilde T \frac{\partial z(\tilde
x)}{\partial \tilde T}} \right|_{\tilde T=\tilde{T}_\textrm{H}^{(0)}}
\label{1loop}
\end{equation}
where $\tilde{T}_\textrm{H}^{(0)}$ is the free Hagedorn temperature of a
specific theory, $z(\tilde x)$ is the corresponding letter partition
function and
\begin{equation}
\langle D_2(\tilde x)\rangle=\sum_{A_1,A_2 \in
\mathcal{A}}\tilde{x}^{d(A_1)+d(A_2)}\langle
A_1A_2|D_2|A_1A_2\rangle \label{d2}
\end{equation}
is the expectation value of the corresponding one-loop dilatation
operator~\cite{Spradlin:2004pp, Harmark:2006di}.

To compute $\langle D_2(\tilde x) \rangle$ in the presence of
chemical potentials for the R-charges and for the Cartan generators
of the $SO(4)$ symmetry we generalize the procedure used
in~\cite{Harmark:2006di}.
In general, $\langle D_2(x,\omega_i,\Omega_i)\rangle$ corresponds to
the expectation value of the one-loop dilatation operator $D_2$
acting on the product of two copies of the singleton representation
$\mathcal{A}\times \mathcal{A}$. From Eq.~\eqref{d2op} we have that
\begin{equation}
\langle D_2(x,\omega_i,\Omega_i)\rangle=
\sum_{j=0}^{\infty}h(j)\frac{V_j(x,\omega_i,\Omega_i)}{\left(1+x^2-2x\cosh(\beta
\omega_1)\right)\left(1+x^2-2x\cosh(\beta \omega_2)\right)}
\label{expD2}
\end{equation}
where $V_j(x,\omega_i,\Omega_i)$ can be computed using the results
presented in~\cite{Harmark:2006di} where in this case we define
\begin{equation}
F_{[k,p,q]}^{(j_L,j_R)} = W_{[k,p,q]} \sum_{m=-j_L}^{j_L} \sum_{\bar
m=-j_R}^{j_R}\rho^m \bar \rho^{\bar m} ,\qquad
W_{[k,p,q]}
\equiv {\tr}_{[k,p,q]} \left( y_i^{J_i} \right),
\end{equation}
with $\rho$ and $\bar \rho$ defined in \eqref{rhobar} and where the
expressions of $W_{[k,p,q]}$ for the various representations are
given in~\cite{Harmark:2006di}.

The general procedure described above allows us to compute
$\langle D_2(x,\omega_i,\Omega_i)\rangle$ for $\CN=4$ SYM on $\R \times S^3$ with all five chemical potentials turned on.
By taking the various decoupling limits we obtain
expressions for $\langle D_2(\tilde x)\rangle$ in each decoupled
theory.\footnote{One can also compute $\langle D_2\rangle$ in the
decoupled theories using the general procedure found in \cite{Bianchi:2006ti}.}

We now have all the ingredients needed to find the one-loop
correction to the Hagedorn temperature from Eq.~\eqref{1loop}. We
end this section with a list of results for the letter partition
function, the expectation value of $D_2$ and the Hagedorn
temperature up to one-loop order for all the non-trivial decoupled theories.
The trivial theories are the bosonic $U(1)$ with $z(\tilde x)=\tilde x$
and the fermionic $U(1)$ with $z(\tilde x)={\tilde x}^{\frac{3}{2}}$.
In both of these theories $D_2$ vanishes and there is no Hagedorn singularity.

\begin{description}
\item[The $SU(2)$ theory]
\begin{equation}
z(\tilde x)=2 \tilde{x}, \quad
\langle D_2(\tilde x)\rangle=\tilde{x}^2, \quad
\tilde{T}_\textrm{H}=
\frac{1}{\log{2}} +\frac{\tilde{\lambda}}{4\log 2} + \CO (\tilde{\lambda}^2)
\end{equation}
\item[The $SU(1|1)$ theory]
\begin{align}
\label{eq:HagSU11}
z(\tilde x)&=\tilde x+{\tilde x}^{\frac{3}{2}}, \quad
\langle D_2(\tilde x)\rangle=\tilde{x}^{\frac{5}{2}}+\tilde{x}^3, \quad
\tilde{T}_\textrm{H}=\tilde{T}_\textrm{H}^{(0)}+\frac{2\, \tilde{T}_\textrm{H}^{(0)} \tilde{\lambda}}{3+2e^{{1}/{2\tilde{T}_\textrm{H}^{(0)}}}} + \CO(\tilde{\lambda}^2), \\
\label{eq:uglyT}
\tilde{T}_\textrm{H}^{(0)}&=
{1}/{\log\left[\frac{1}{3}-\frac{5}{3}\left(\frac{2}{11+3\sqrt{69}}\right)^{1/3}+
\frac{1}{6}\left(11+3\sqrt{69}\right)^{1/3}\right]}
\end{align}
\item[The $SU(1|2)$ theory]
\begin{equation}
z(\tilde x)=2\tilde x+{\tilde x}^{\frac{3}{2}},\quad
\langle D_2(\tilde x)\rangle=\left(\tilde x+\tilde{x}^{\frac{3}{2}}\right)^2, \quad
\tilde{T}_\textrm{H}=\frac{1}{\log{\frac{2}{3-\sqrt
5}}}\left(1+ \frac{4 }{5+3\sqrt 5} \tilde{\lambda}+ \CO
(\tilde{\lambda}^2) \right)
\end{equation}
\item[The $SU(2|3)$ theory]
\begin{equation}
z(\tilde x)=3\tilde x+2{\tilde x}^{\frac{3}{2}}, \quad
\langle D_2(\tilde x)\rangle=3\tilde{x}^2+6\tilde{x}^{\frac{5}{2}}+3\tilde{x}^3, \quad
\tilde{T}_\textrm{H}=\frac{1}{\log{4}}\left(1+\frac{3}{8} \tilde{\lambda}+ \CO (\tilde{\lambda}^2) \right)
\end{equation}
\item[The bosonic $SU(1,1)$ theory]
\begin{equation}
z(\tilde x)=\frac{\tilde x}{1-\tilde x}, \quad
\langle D_2(\tilde x)\rangle=-\frac{\tilde{x}^2 \log(1-\tilde{x})}{(1-\tilde{x}^2)^2},\quad
\tilde{T}_\textrm{H}=\frac{1}{\log{2}}+\frac{1}{2} \tilde{\lambda}+ \CO (\tilde{\lambda}^2)
\end{equation}
\item[The fermionic $SU(1,1)$ theory]
\begin{equation}
z(\tilde x)=\frac{{\tilde x}^{\frac{3}{2}}}{1-\tilde x}, \quad
\langle D_2(\tilde x)\rangle=-\frac{\tilde{x}^2 \log(1-\tilde{x})}{(1-\tilde{x}^2)^2},\quad
\tilde{T}_\textrm{H}=\tilde{T}_\textrm{H}^{(0)}
+\frac{3 e^{{1}/{2 \tilde{T}_\textrm{H}^{(0)}}}\, \tilde{\lambda}}{3 e^{{1}/{\tilde{T}_\textrm{H}^{(0)}}}-1}
+ \CO(\tilde{\lambda}^2),
\end{equation}
with $\tilde{T}_\textrm{H}^{(0)}$ the same as in Eq.~\eqref{eq:uglyT}.
\item[The $SU(1,1|1)$ theory]
\begin{equation}
z(\tilde x)=\frac{\tilde x}{1-\sqrt{\tilde x}},\quad
\langle D_2(\tilde x)\rangle=
-\frac{\tilde{x}^{\frac{3}{2}} \log(1-\tilde{x})}{(1-\sqrt{\tilde{x}})^2},\quad
\tilde{T}_\textrm{H}=\frac{1}{\log{\frac{2}{3-\sqrt{5}}}}+ \frac{1}{\sqrt 5}\tilde{\lambda}
+ \CO (\tilde{\lambda}^2)
\end{equation}
\item[The $SU(1,1|2)$ theory]
\begin{equation}
z(\tilde x)=\frac{2\tilde x}{1-\sqrt{\tilde x}},\quad
\langle D_2(\tilde x)\rangle=
-\frac{\tilde{x}(1+\sqrt{\tilde{x}})^2 \log(1-\tilde{x})}{(1-\sqrt{\tilde{x}})^2},\quad
\tilde{T}_\textrm{H}=\frac{1}{2\log{2}}+ \frac{3\log{\frac{4}{3}}}{4\log 2}
\tilde{\lambda}+ \CO (\tilde{\lambda}^2)
\end{equation}
\item[The $SU(1,2)$ theory]
\begin{equation}
z(\tilde x)=\frac{{\tilde x}^2}{(1-\tilde x)^2},\quad
\langle D_2(\tilde x)\rangle=\frac{\tilde{x}^3+(\tilde{x}^2-2{\tilde{x}^3}) \log
(1-\tilde{x})}{(1-{\tilde{x}})^4},\quad
\tilde{T}_\textrm{H}=\frac{1}{\log{2}}+
\frac{\tilde{\lambda}}{2\log 2}+ \CO (\tilde{\lambda}^2)
\end{equation}
\item[The $SU(1,2|1)$ theory]
\begin{align}
z(\tilde x)= & \frac{{\tilde x}^{\frac{3}{2}}}{(1-\sqrt{\tilde x})^2(1+\sqrt{\tilde x})},\quad
\langle D_2(\tilde x)\rangle=
\frac{\tilde{x}^{\frac{5}{2}}-(\tilde{x}^{\frac{5}{2}}+\tilde{x}^{2}-\tilde{x}^{\frac{3}{2}}) \log
(1-\tilde{x})}{(1-\sqrt{\tilde{x}})^2(1-{\tilde{x}})^2}, \nn\\
\tilde{T}_\textrm{H}=&\frac{1}{\log{\frac{2}{3-\sqrt{5}}}}\left(1+ \sqrt{\frac{2}{5} (3-\sqrt 5)}\,
\tilde{\lambda}+ \CO (\tilde{\lambda}^2) \right)
\end{align}
\item[The $SU(1,2|2)$ theory]
\begin{equation}
z(\tilde x)=\frac{\tilde x}{(1-\sqrt{\tilde x})^2},\quad
\langle D_2(\tilde x)\rangle=
\frac{\tilde x^2+(\tilde{x}-2\tilde{x}^{\frac{3}{2}}) \log (1-\tilde{x})}%
{(1-\sqrt{\tilde{x}})^4},\quad
\tilde{T}_\textrm{H}=\frac{1}{2\log{2}} + \frac{\tilde{\lambda}}{4\log 2} +
\CO (\tilde{\lambda}^2)
\end{equation}
\item[The $SU(1,2|3)$ theory]
\begin{align}
z(\tilde x)=&\frac{3 \tilde x - \tilde x ^{\frac{3}{2}}}{(1-\sqrt{\tilde x})^2}, \\
\langle D_2(\tilde x)\rangle
=&\frac{\tilde x^{\frac{3}{2}}}{(1-{\tilde{x}})^4}
\left[\left(1+6{\tilde x}^\frac{1}{2}+15\tilde x+20\tilde{x}^{\frac{3}{2}}
       +21\tilde x^2+6\tilde{x}^{\frac{5}{2}}-19\tilde x^3+10\tilde x^4\right)\right. \nn\\
&+\left.\left(1+3\tilde x^\frac{1}{2}-2\tilde x-19\tilde{x}^{\frac{3}{2}}
       -24\tilde x^2-19\tilde{x}^{\frac{5}{2}}-4\tilde x^3+3\tilde{x}^{\frac{7}{2}}
       +\tilde x^4\right)\log(1-\tilde{x})\right], \nn\\
\tilde{T}_\textrm{H}=&\frac{1}{\log{\frac{2}{7-3\sqrt{5}}}}
\left(1-
\frac{16[201341-90043\sqrt 5+(262\sqrt 5-586)\log{\frac{3\sqrt{5}-5}{2}}]}%
{5(5-3\sqrt 5)^4}
\tilde{\lambda}
+ \CO (\tilde{\lambda}^2) \right)\nn
\end{align}
\end{description}


\subsection{Large $\tilde{\lambda}$ limit of theories containing scalars}
\label{sec:largelambda}

In this section we use the low energy spectrum
\eqref{eq:ABCspectrum}--\eqref{eq:ABCconstraint} obtained from the
general Bethe ansatz in the thermodynamic limit to study the large
$\tilde\lambda$ Hagedorn temperature for the nine decoupled theories that
contain scalars. The three remaining theories are discussed in
Section~\ref{sec:largelambsu12}.
This limit of the Hagedorn temperature has been calculated
for the decoupled $SU(2)$ theory in \cite{Harmark:2006ta} and for the
$SU(2)$ theory coupled to a magnetic field in \cite{Harmark:2006ie}.
We will use the same methods here to obtain a general expression for the
large $\tilde\lambda$ Hagedorn temperature that is valid for all the nine
non-trivial theories that contain at least one scalar. The Hagedorn
temperature will depend on the numbers given in Table~\ref{tab:abc}.

There is a direct connection between the Hagedorn temperature in the
decoupled theories and the free energy per site of the corresponding spin
chain model in the thermodynamic limit \cite{Harmark:2006ta}.
For all the decoupled theories that contain a scalar we consider the function
\begin{align}
\label{eq:defV}
V(\tilde\beta) \equiv \lim_{L\to\infty} \frac{1}{L}\log\left[
{\tr}_L \left( e^{-\tilde\beta (H-L)}\right) \right].
\end{align}
The limit is finite since $V(\tilde\beta)$ is related to the
thermodynamic limit of the free energy per site $f$ by
$V(\tilde\beta)=-\tilde\beta f(\tilde\beta )$. The spectrum of the
Hamiltonian is given in Eq.~\eqref{eq:ABCspectrum}. Note that in the
definition of $V(\tilde \beta)$ we subtract from the Hamiltonian the
constant contribution $L$ coming from $D_0$ but include the other
contributions from $D_0$ that depend on the state of the spin chain.
If we view $\tilde\lambda D_2$ as the Hamiltonian of the spin chain,
then these additional terms from $D_0$ can in most cases be viewed
as a coupling to an external magnetic field as in  \cite{Harmark:2006ie}.

For all of the decoupled theories the partition function is given by
\begin{align}
\log Z(\tilde\beta) = \sum_{n=1}^{\infty}\sum_{L=1}^{\infty}
\frac{1}{n}e^{-n \tilde\beta L} {\tr}_L \left( e^{-n\tilde\beta (H-L)}\right)
\end{align}
for any value of $\tilde{\lambda}$. For large $L$ we have that
\begin{align}
e^{-n \tilde\beta L} {\tr}_L \left( e^{-n\tilde\beta (H-L)}\right) \simeq
\exp\left(-n \tilde\beta L + L V(n \tilde\beta)\right).
\end{align}
From this observation we see that the Hagedorn temperature $\tilde{T}_{\rm H} = 1/\tilde{\beta}_{\rm H}$ for any of value of $\tilde{\lambda}$ is determined by the equation \cite{Harmark:2006ta}
\begin{align}
\label{eq:hagedorncondition}
\tilde\beta_\textrm{H} = V(\tilde\beta_\textrm{H}).
\end{align}
We use this general equation for the Hagedorn temperature below to find the Hagedorn temperature for large $\tilde\lambda$. In
Section~\ref{sec:SU11asXX} we give an exact expression for
$V(\tilde\beta)$ for the $SU(1|1)$ theory and use that to obtain the
Hagedorn temperature for small $\tilde\lambda$ in that case as well.

In the following we
use our knowledge of the low energy spectrum \eqref{eq:ABCspectrum}
to obtain the Hagedorn temperature for large $\tilde\lambda$.
Recall that the low energy spectrum can be written in the form of
Eq.~\eqref{eq:ABCspectrum} using the number operators $M_n, N_n$,
and $F_n$.  In order to find $V(\tilde\beta)$ we are interested in
the large $L$ behavior of
\begin{align}
{\tr}_L \left( e^{-\tilde\beta (H - L)}\right)
&= \sum_{\{M_n\}}\sum_{\{N_n\}}\sum_{\{F_n\}}
\int_{-\frac{1}{2}}^{\frac{1}{2}} du\,  \exp \left\{ -\tilde\beta \left(H-L\right)
 + 2\pi i u P  \right\}
\end{align}
where $H-L$ is given by Eq.~\eqref{eq:ABCspectrum}
and the integration over $u$ has been introduced to impose the
zero momentum constraint in the spectrum \eqref{eq:ABCconstraint}.
Evaluating the sums over the number operators with $M_n$ and $N_n$
ranging from zero to infinity and $F_n$ from zero to one, we get
\begin{align}
 {\tr}_L \left( e^{-\tilde\beta (H - L)}\right)
&= \int_{-\frac{1}{2}}^{\frac{1}{2}} \!\!\! du\  \prod_{n \in \mathbb{Z}}
\frac{
  \left(
    1+\exp\left(
      \frac{2\pi^2\tilde\beta\tilde\lambda}{L^2}n^2
      -\frac{\tilde\beta}{2} + 2\pi i u n
     \right)\right)^c
}{
  \left(
    1-\exp\left(
      \frac{2\pi^2\tilde\beta\tilde\lambda}{L^2}n^2 + 2\pi i u n
    \right)\right)^a
  \left(
    1-\exp\left(
      \frac{2\pi^2\tilde\beta\tilde\lambda}{L^2}n^2
      - \tilde\beta + 2\pi i u n
    \right)\right)^b}.
\end{align}
Analysis similar to the one in \cite{Harmark:2006ta} shows that
the leading contribution
for  $L\gg 1$ comes from $u=0$ and that it is given by
\begin{align}
{\tr}_L \left( e^{-\tilde\beta (H-L)}\right) \sim
\exp \left(
\frac{L}{\sqrt{2\pi\tilde\lambda\tilde\beta}}
\sum_{p=1}^{\infty} \frac{a + b \left (e^{-\tilde\beta}\right)^p - c \left(-e^{ \tilde\beta/2 }\right)^{p}}{p^{3/2}}
\right).
\end{align}
Using this in Eq.~\eqref{eq:defV} we arrive at
\begin{align}
\label{eq:ABCV}
V(\tilde\beta) = \frac{1}{\sqrt{2\pi\tilde\beta\tilde\lambda}}\left[
a\, \zeta(3/2)
+ b \, \mathrm{Li}_{3/2}\left(e^{-\tilde\beta}\right)
- c \, \mathrm{Li}_{3/2}\left(-e^{-\tilde\beta/2}\right)
\right]
\end{align}
where $\zeta(x)$ is the Riemann zeta function and
$\mathrm{Li}_n(x)$ is the Polylogarithm function.
We can now solve the equation $V(\tilde\beta_\textrm{H})=\tilde\beta_\textrm{H}$
to get the Hagedorn temperature for large $\tilde\lambda$
\begin{align}
\label{eq:ABChage}
\tilde T_\textrm{H} = \left(\frac{2\sqrt{2\pi}}{(2a + 2b + (2-\sqrt{2})c)\, \zeta(3/2)}\right)^{2/3}
\tilde\lambda^{1/3}
\end{align}
This expression is valid for all theories that contain at least one scalar. The numbers
$a,b,c$ for each such theory are given in Table~\ref{tab:abc}. Note that \eqref{eq:ABChage} correctly reduces to the result obtained for the $SU(2)$ decoupled theory in \cite{Harmark:2006ta} for $a=1$ and $b=c=0$.

\subsection{The $SU(1|1)$ theory as a magnetic $XX$ Heisenberg spin chain}
\label{sec:SU11asXX}

In this section we rewrite the Hamiltonian of the decoupled
$SU(1|1)$ theory as a Heisenberg $XX_{1/2}$ spin chain coupled to an
external magnetic field.  The spin chain model is exactly solvable
and using known results on the free energy we can in principle
obtain the Hagedorn temperature for any value of $\tilde\lambda$. We
demonstrate how the Hagedorn temperature can be obtained to
arbitrary order in small $\tilde\lambda$ and for large
$\tilde\lambda$ we verify that the exact result agrees with our
Bethe ansatz method to obtain $V(\tilde\beta)$.

\begin{figure}[ht]
\centerline{\epsfig{file=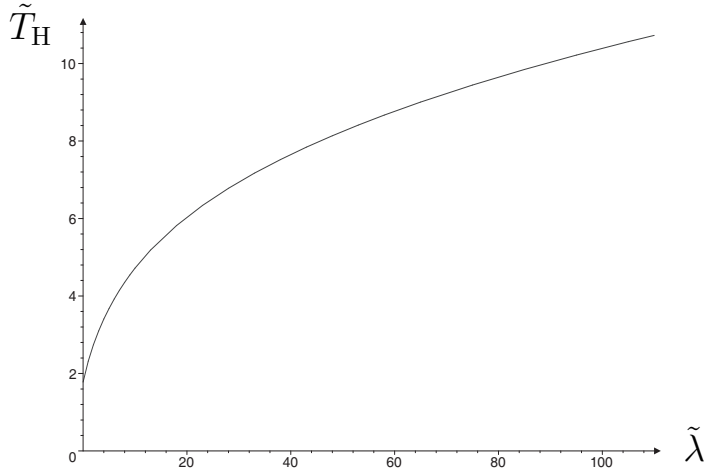,width=8cm,height=6cm} }
\caption{Hagedorn temperature $\tilde{T}_{\rm H}$ as a function of
$\tilde{\lambda}$ for the $SU(1|1)$ decoupled
theory.\label{fig:SU11}}
\begin{picture}(0,0)(0,0)
\put(87,205){\Large $\tilde{T}_{\rm H}$}
\put(342,45){\Large $\tilde{\lambda}$}
\end{picture}
\end{figure}

Following \cite{Staudacher:2004tk}, we rewrite the
$D_2$ part of the $SU(1|1)$ Hamiltonian in spin chain form by
expressing it in terms of the three Pauli matrices
$\sigma^1,\sigma^2, \sigma^3$ as
\begin{align}
\label{xy}
D_2 = \frac{1}{2}\sum_{j = 1}^L (1-\Pi_{j,j+1})
= \sum_{j=1}^L\,\frac{1}{2}\left(
(1-\sigma_j^3)
-\frac{1}{2} (\sigma_j^1 \sigma_{j+1}^1+\sigma_j^2 \sigma_{j+1}^2)
\right) .
\end{align}
In spin chain language the bosonic partons $Z$ are spin-up
spinors and the fermionic partons $\chi_1$ are spin-down.
The $D_0$ part of the Hamiltonian can similarly be expressed in terms
of the Pauli matrices as
\begin{align}
D_0 = L + \sum_{j=1}^L \frac{1}{4} (1-\sigma_j^3)
\end{align}
and the full decoupled $SU(1|1)$ Hamiltonian can therefore
be written as
\begin{align}
H_{SU(1|1)} = L + \sum_{j=1}^L \left(
\frac{1}{4}(1+2\tilde\lambda)\left(1-\sigma_j^3\right)
- \frac{\tilde\lambda}{4}  \left(\sigma_j^1 \sigma_{j+1}^1+\sigma_j^2 \sigma_{j+1}^2\right)
\right)
\end{align}
which is the Heisenberg $XX_{1/2}$ spin chain Hamiltonian with
nearest neighbor coupling $\tilde\lambda/2$ in an external magnetic
field of strength $(1+2\tilde\lambda)/4$. This spin chain is exactly
solvable and an expression for the free energy per site which is
valid for all values of $\tilde\lambda$ is known
\cite{PhysRev.127.1508}. In our notation this translates into
\begin{align}
\label{eq:Vfor1slash1}
V(\tilde\beta)= -\frac{\tilde\beta}{4}(1+2\tilde\lambda)
+ \frac{1}{\pi}\int_0^\pi d\omega \log
    \left[2\cosh\left(\frac{\tilde\beta}{2}\left\{\frac{1}{2}+\tilde\lambda(1-\cos\omega)\right\}\right)\right].
\end{align}
From this function we can obtain the Hagedorn temperature for all
values of $\tilde\lambda$ by employing the general equation
\eqref{eq:hagedorncondition} for the Hagedorn temperature. We have
used this to plot the Hagedorn temperature $\tilde{T}_{\rm H}$ as a
function of $\tilde{\lambda}$ in Figure \ref{fig:SU11}.

\subsubsection*{Hagedorn temperature for small $\tilde\lambda$}

Let us first verify that can we match the $\tilde\lambda\to 0$ limit of the
above considerations with the free Hagedorn temperature computation in
Section~\ref{sec:freetheory}.  We immediately get the condition
\begin{align}
\tilde\beta_\textrm{H} = V(\tilde\beta_\textrm{H})\large\vert_{\tilde\lambda=0} =
-\frac{\tilde\beta_\textrm{H}}{4}  + \log \left(2\cosh \frac{\tilde\beta_\textrm{H}}{4}\right)
\end{align}
which is equivalent to
\begin{align}
e^{-\tilde\beta_\textrm{H}} +  e^{-3\tilde\beta_\textrm{H}/2} = 1.
\end{align}
This is the same equation as obtained from the free letter partition
function of the $SU(1|1)$ theory in Secton~\ref{sec:smalllambda} and
the free Hagedorn temperature $\tilde T_\textrm{H}^{(0)}$ is given
in Eq.~\eqref{eq:uglyT}. Equipped with the exact
expression for $V(\tilde\beta)$ we can now go further and obtain
higher loop corrections for the Hagedorn temperature. Expanding
Eq.~\eqref{eq:Vfor1slash1} for small $\tilde\lambda$ yields
\begin{align}
V(\tilde\beta) =
-\frac{\tilde\beta }{4}
+\log 2\cosh \left({\tilde\beta}/{4}\right)
+\frac{\tilde\lambda \tilde\beta}{2} \left(\tanh(\tilde\beta/4)-1\right)
+\frac{3 \tilde\beta^2 \tilde\lambda ^2}{16}  \text{sech}^2\left(\tilde\beta /4\right)
+O\left(\tilde\lambda^3\right)
\end{align}
and solving the equation $\tilde\beta_\textrm{H} = V(\tilde\beta_\textrm{H})$
then gives the Hagedorn temperature to this order
\begin{align}
\tilde T_\textrm{H} = \tilde T_\textrm{H}^{(0)} +
\frac{2 \tilde T_\textrm{H}^{(0)}  \tilde\lambda }{3+2 e^{1 /2\tilde T_\textrm{H}^{(0)}}}
-\frac{e^{b /2}\left(17+28 e^{b /2}+12 e^{b }\right)\tilde\lambda ^2}{
2\left(1+e^{b /2}\right) \left(3+2 e^{b/2}\right)^3} +O\left(\tilde\lambda ^3\right)
\end{align}
where we have introduced the short hand notation $b=1/\tilde
T_\textrm{H}^{(0)}$ to simplify the two-loop term. It is a
comforting check that the one-loop term is precisely the same as
found in Eq.~\eqref{eq:HagSU11}.  Using the spin chain method we can
easily obtain $\tilde T_\textrm{H}$ to arbitrarily high order in
$\tilde\lambda$.

\subsubsection*{Hagedorn temperature for large $\tilde\lambda$}

From Eq.~\eqref{eq:ABCV} we already know the leading behavior of
$V(\tilde\beta)$ for large $\tilde\lambda$ and large $L$. As a check
of that result we can extract the large $\tilde\lambda$ behavior of the
exact function $V(\tilde\beta)$ in Eq.~\eqref{eq:Vfor1slash1} and compare
the two. From Eq.~\eqref{eq:ABChage} we know that
$\tilde\beta_\textrm{H} \sim \tilde\lambda^{-1/3}$ for large $\tilde\lambda$
and we are therefore interested in large $\tilde\lambda\tilde\beta$.
In this limit we find
\begin{align}
V(\tilde\beta) \simeq \frac{1}{\pi} \int_{0}^\pi d\omega
\log\left[1+\exp\left(-\tilde\lambda\tilde\beta(1-\cos\omega)\right)\right]
\simeq \frac{(\sqrt{2}-1)\zeta(3/2)}{\sqrt{4\pi \tilde\lambda\tilde\beta}}.
\end{align}
The leading contribution comes from integrating over small $\omega$
and we therefore used the saddle-point approximation to get the final
result.  This is the same expression as Eq.~\eqref{eq:ABCV}
with $a=b=0$, $c=1$, and the polylogarithm expanded for $\tilde\beta \ll 1$.


\subsection{Large $\tilde{\lambda}$ limit of the $SU(1,2)$ theory}
\label{sec:largelambsu12}

In Section \ref{sec:largelambda} we found the large
$\tilde{\lambda}$ behavior of the Hagedorn temperature $\tilde{T}_\textrm{H}$
in the decoupled theories containing scalars. This was done by considering
the low energy behavior of the spin chain with Hamiltonian $H-L$,
where $H = D_0+\tilde{\lambda} D_2$. In the following we shall see
that in theories without scalars the low energy behavior of the spin chain
Hamiltonian cannot be connected with the large $\tilde{\lambda}$ behavior.
We illustrate this by considering the
$SU(1,2)$ theory, which is particularly interesting since the
decoupled states are states of pure Yang-Mills
theory, as we explore further in Section \ref{sec:pureYM}. We
comment below on the consequence of our observations for obtaining a
string dual of the $SU(1,2)$ theory.

For the $SU(1,2)$ we found in Section \ref{sec:lowspec} the spectrum
\eqref{eq:Hsu12}. This spectrum is accurate in the limit when $L$ is
large and $\tilde{\lambda}$ is large, and for states with
$H-(2+3\tilde{\lambda}/2)L$ not large. In analogy with
\eqref{eq:defV} we define the function $V_{SU(1,2)}(\tilde{\beta})$
as
\begin{align}
\label{Vsu12} V_{SU(1,2)}(\tilde\beta) \equiv \lim_{L\to\infty}
\frac{1}{L}\log\left[ {\tr}_L \left( e^{-\tilde\beta
(H-(2+\frac{3}{2} \tilde{\lambda}) L)}\right) \right].
\end{align}
Consider then the large $\tilde{\lambda}$ limit of \eqref{Vsu12}. Using \eqref{eq:Hsu12} we see that the
spectrum of $H-(2+\frac{3}{2} \tilde{\lambda}) L$
consists of a magnetic part and a
part which is proportional to $\tilde{\lambda}/L^2$. While the
magnetic part does not receive finite size correction, the other
part does. From this, one sees that the
finite size corrections are suppressed in
$V_{SU(1,2)}(\tilde{\beta})$ provided that $\tilde{\lambda}
\tilde{\beta} \gg 1$. Therefore, we can use the spectrum
\eqref{eq:Hsu12} to find that
\begin{equation}
\label{appVsu12} V_{SU(1,2)}(\tilde\beta) \simeq \frac{\sqrt
6}{\sqrt{\pi \tilde\lambda\tilde\beta}}
\Li_{3/2}\left(e^{-\tilde\beta}\right) \ \ \mbox{for} \ \
\tilde{\lambda} \tilde{\beta} \gg 1.
\end{equation}

Now, consider the Hagedorn temperature for the $SU(1,2)$ theory in
general. Following the argument of Section \ref{sec:largelambda} we
see that the Hagedorn temperature $\tilde{T}_\textrm{H}=1/\tilde{\beta}_\textrm{H}$
for any value of $\tilde{\lambda}$ is given by the equation
\begin{equation}
\label{su12hag} \left( 2+ \frac{3}{2} \tilde{\lambda} \right)
\tilde{\beta}_\textrm{H} = V_{SU(1,2)}(\tilde{\beta}_\textrm{H}).
\end{equation}
Take then the large $\tilde{\lambda}$ limit. We see first that
Eq.~\eqref{su12hag} becomes $3 \tilde{\lambda} \tilde{\beta}_\textrm{H}
\simeq 2 V_{SU(1,2)}(\tilde{\beta}_\textrm{H})$. Then, if we try to insert
the approximation \eqref{appVsu12} for $V_{SU(1,2)}(\tilde{\beta})$
we get the equation
\begin{equation}
\label{badrel} \frac{3\sqrt{\pi}}{2\sqrt{6}} \big(\tilde{\lambda}
\tilde{\beta}_\textrm{H} \big)^{\frac{3}{2}} \simeq
\Li_{3/2}\left(e^{-\tilde\beta_\textrm{H}}\right).
\end{equation}
However, $\Li_{3/2} ( e^{-x} )$ is a decreasing function of $x$
bounded from above by $\Li_{3/2} (1) = \zeta(3/2)$. Thus, the
relation \eqref{badrel} requires that $\tilde{\lambda}
\tilde{\beta}_\textrm{H}$ is of order one or smaller. Clearly, this conflicts
with the approximation used to derive \eqref{appVsu12}. Therefore,
we cannot infer the behavior of the Hagedorn temperature
$\tilde{T}_\textrm{H}$ for large $\tilde{\lambda}$ using the result
\eqref{appVsu12}.

We can therefore conclude that the free magnon spectrum \eqref{eq:Hsu12}
does not correspond to the behavior of the $SU(1,2)$ theory for
large $\tilde{\lambda}$. To understand the large $\tilde{\lambda}$
behavior of the $SU(1,2)$ one must therefore solve the full Bethe
equations \eqref{eq:genBethe1}--\eqref{eq:genBethe2} for that theory.
Therefore, contrary to the theories with scalars, the large
$\tilde{\lambda}$ limit does not correspond to a free magnon limit of the
spin chain for this decoupled theory.

That we cannot use the free spectrum \eqref{eq:Hsu12} to approximate
large $\tilde{\lambda}$ for the $SU(1,2)$ theory means that it is
considerably harder to understand the $SU(1,2)$ decoupling limit on
the string theory side in the AdS/CFT correspondence. In
\cite{Harmark:2006ta} it was found for the $SU(2)$ theory how to
obtain the spectrum and the Hagedorn temperature for large
$\tilde{\lambda}$ from the string side. However, it is not clear how
to find a similar match of the spectrum and Hagedorn temperature for
the $SU(1,2)$ theory since it is not well understood how to obtain
the full set of finite-size effects on the string side.

Note finally that the above considerations for the $SU(1,2)$ theory
can be repeated for the other two decoupled theories without
scalars, $i.e.$ the fermionic $SU(1,1)$ and the $SU(1,2|1)$
theories, with analogous results. Thus, also for these two theories
the large $\tilde{\lambda}$ behavior is not linked to the free
magnon limit of a spin chain.

\section{Microcanonical version of the decoupling limits}
\label{sec:micro}

The decoupling limits described in Section \ref{sec:newlim} are
taken of the partition function of $\CN=4$ SYM on $\R \times S^3$ in
the grand canonical ensemble. It is highly useful to understand how
the decoupling limits are taken in the microcanonical ensemble. In
particular, this is necessary in order to translate these decoupling
limits to the string side of the AdS/CFT correspondence.
In Section \ref{sec:microlim} we consider
how to implement the decoupling limits in the microcanonical
ensemble and in Section
\ref{sec:stringreg} we use the microcanonical decoupling limits to
identify, for any of the decoupling limits containing scalars, a
regime of weakly coupled planar $\CN=4$ SYM on $\R \times S^3$ in
which it corresponds to tree-level string theory.

Another reason why it is important to obtain an understanding of
our decoupling limits in the microcanonical ensemble is that one can
think of the decoupling limits solely in terms of gauge invariant
operators of $\CN=4$ SYM on $\R^4$. This is in contrast to the grand
canonical ensemble in which the correct interpretation is rather in
terms of the partition function which sums over states of $\CN=4$
SYM on $\R \times S^3$. Thus, both the microcanonical decoupling
limits that we present below in Section \ref{sec:microlim} and the
``stringy regime" that we present in Section \ref{sec:stringreg}
apply also to gauge-invariant operators of $\CN=4$ SYM on $\R^4$.

\subsection{Microcanonical limit}
\label{sec:microlim}

Let $n=(n_1,n_2,n_3,n_4,n_5)$ be given such that it fulfils the
requirements described in Section \ref{sec:gencon}. Let $J$ be
defined as in \eqref{defJ}. Then the decoupling limit of $SU(N)$
$\CN=4$ SYM on $\R \times S^3$ in the microcanonical ensemble is
given as
\begin{equation}
\label{miclim} \lambda \rightarrow 0, \quad
\tilde{H} \equiv \frac{D-J}{\lambda} \ \mbox{fixed}, \quad
J, N \ \mbox{fixed}
\end{equation}
where $D$ is the dilation operator which is expanded as \eqref{dilop}
for small $\lambda$. We see that the limit \eqref{miclim}
indeed is in the microcanonical ensemble since $\tilde{H}$ and $J$
are linear combinations of the Cartan generators of $psu(2,2|4)$.
Analyzing the limit \eqref{miclim} we see that since $D \simeq D_0 +
\lambda D_2$ for small $\lambda$ only states with $D_0 = J$ survive
and we get that $\tilde{H} = D_2$ for $D_2$ acting on the surviving
states.

We first observe that the set of states/operators that we have after
the decoupling limit are the ones with $D_0 = J$ for a given $J$.
Thus, whereas for the grand canonical limit \eqref{declim} we had
all states with $D_0 = J$ for any choice of $J$, for the
microcanonical limit we only have the subset of states corresponding
to a particular fixed value of $J$. Therefore, the microcanonical
decoupling limit \eqref{miclim} is seen to give a subset of the
decoupled states that we get in the grand canonical decoupling limit
\eqref{declim}.

We furthermore observe that while for the grand canonical limit
\eqref{declim} we have $D_0+\tilde{\lambda} D_2$ as the effective
Hamiltonian, we have $\tilde{H} = D_2$ as the effective Hamiltonian
for the microcanonical limit \eqref{miclim}. This is in accordance
with the fact that we pick a fixed $J$ in the decoupled theory,
since we clearly have that the $D_0 + \tilde{\lambda} D_2$ Hamiltonian
is equivalent to choosing $D_2$ as the Hamiltonian if we keep $D_0$
fixed. We can therefore conclude that the two decoupling limits
\eqref{declim} and \eqref{miclim} give us the same decoupled theory
in two different ensembles, $i.e.$  in the \eqref{declim} limit we
end up in the canonical ensemble while in the \eqref{miclim} limit
we end up in the microcanonical ensemble.

Another important point is that it follows from the commutation
relation $[D_2,D_0]=0$ in Eq.~\eqref{D2coms} that $\tilde{H}$ commutes
with $J$. This means that there are no interactions between states
with different values of $J$, $i.e.$ the subsector of the decoupled
theory that we choose by fixing $J$ is closed with respect to
$\tilde{H}$.

We can thus conclude that the microcanonical decoupling limit
\eqref{miclim} for a given $n=(n_1,n_2,n_3,n_4,n_5)$ leads to the
same decoupled theory as the grand canonical limit \eqref{declim}.
It is moreover clear that the same analysis applies concerning which
decoupling limits one has, thus the list of decoupling limits
of Section \ref{sec:declist} applies equally well to the
microcanonical decoupling limit \eqref{miclim}.

In the planar limit $N = \infty$ of $\CN=4$ SYM on $\R \times S^3$
and in the decoupling limit \eqref{miclim}, with
$n=(n_1,n_2,n_3,n_4,n_5)$ chosen from the list in Section
\ref{sec:declist}, the spectrum of $\tilde{H}=D_2$ is given by
\eqref{eq:genBetheSpectrum} with the Bethe roots determined by
\eqref{eq:genBethe1}--\eqref{eq:genBethe2}. Thus, we have the full
spectrum for the decoupled theory in the planar limit and each of
the limits of Section \ref{sec:declist} correspond to an integrable
spin chain.

However, there is a subtle issue in applying the spin chain picture
to the microcanonical decoupling limit \eqref{miclim}. In the
microcanonical limit \eqref{miclim} we fix $J$, whereas when
applying the Bethe equation \eqref{eq:genBethe1} we consider a
certain length $L$ of the spin chain. However, in general the length
$L$ is not fixed for a given $J$. For instance, in the bosonic
$SU(1,1)$ limit the operators $\tr ( Z Z Z )$ and $\tr ( Z d_1 Z )$
both have $J=3$ while $L$ is 3 and 2, respectively. Therefore, when
applying the Bethe ansatz technique, one should divide the decoupled
theory into the different subsectors according to the possible
values of $L$, and then apply the Bethe ansatz technique separately
for these subsectors. It is however necessary for this to work that
there are no interactions between the subsectors of different
lengths. That this is the case can be seen by the fact that the
$D_2$ operator cannot change the length of a state. One way to see
this is to observe that the length operator is $L = 1 - C$, where
$C$ is the central charge of the $u(2,2|4)$ algebra, as reviewed in
Appendix \ref{app:oscrep}. From this fact it is easy to check that
one has $[D_2, L]=0$, hence $D_2$ does not change the length.

\subsection{Regimes of $\CN=4$ SYM with stringy behavior}
\label{sec:stringreg}

In \cite{Harmark:2006ta} it was found that for the $SU(2)$
decoupling limit (see Section \ref{sec:declist}) one can match the
spectrum and Hagedorn temperature as found from the gauge theory and
string theory sides when $\tilde{\lambda} \rightarrow \infty$.

One of the reasons behind the successful match of
\cite{Harmark:2006ta} is that $\tilde{\lambda} = \lambda/(1-\Omega)$
works as an effective 't Hooft coupling in the decoupled theory.
This can for example be seen from the fact that the
$\tilde{\lambda}^n$ contribution to the Hagedorn temperature
$\tilde{T}_\textrm{H}$ origins from part of the $n$-loop diagram for $\CN=4$
SYM on $\R \times S^3$. Therefore, taking the large
$\tilde{\lambda}$ limit can be seen as taking the strong coupling
limit. However, since the effective Hamiltonian is $D_0 +
\tilde{\lambda} D_2$ this can be accomplished in a controllable
manner. Thus, in this sense one can say that the successful match
between gauge theory and string theory in \cite{Harmark:2006ta} is
due to the fact that it was found how to take the strong coupling
limit in a controllable way.

We expect that the match between gauge theory and string theory in
the large $\tilde{\lambda}$ limit works for all the 9 theories in
the list of limits in Section \ref{sec:declist} that include
scalars, $i.e.$  the $SU(2)$, $SU(1|1)$, $SU(1|2)$, $SU(2|3)$,
bosonic $SU(1,1)$, $SU(1,1|1)$, $SU(1,1|2)$, $SU(1,2|2)$ and
$SU(1,2|3)$ limits. This is in accordance with the results of
Section \ref{sec:lowspec} where it is found that for large
$\tilde{\lambda}$ the spectrum is string-like.

The question of this section is then in which regime of planar
$\CN=4$ SYM on $\R \times S^3$ do we see stringy behavior, given any
of these 9 decoupling limits with scalars, $i.e.$  how do we
translate the large $\tilde{\lambda}$ limit, which makes sense in
the grand canonical ensemble, to a statement about $\CN=4$ SYM on
$\R \times S^3$ in the microcanonical ensemble.

Now, taking the large $\tilde{\lambda}$ limit for the effective
Hamiltonian $D_0 + \tilde{\lambda} D_2$ corresponds to considering
the low energy states for the $D_2$ operator (for the decoupled
theories with scalars). In other words, for large $\tilde{\lambda}$
we consider states with $D_2$ of order $1/\tilde{\lambda}$ so that
$\tilde{\lambda} D_2$ is of order one. Since $(D-J)/\lambda$
approaches $D_2$ in the limit \eqref{miclim} we see that we should
have $(D-J)/\lambda$ to be of order $1/\tilde{\lambda}$. Thus, we
need that $|D-J| \ll \lambda$. The limit \eqref{miclim} also
requires $\lambda \ll 1$ and $|D-J| \ll 1$, and in addition we need
large $J$ to see string-like states, so combining these ingredients
we get that the large $\tilde{\lambda}$ limit corresponds to probing
the regime
\begin{equation}
\label{stringregime} | D-J | \ll \lambda \ll 1, \quad J \gg 1.
\end{equation}
Thus, for $n=(n_1,n_2,n_3,n_4,n_5)$ corresponding to one of the nine non-trivial decoupling
limits with scalars listed in Section \ref{sec:declist}, we have identified
the regime \eqref{stringregime} for which planar $\CN=4$ SYM on $\R
\times S^3$ has a string-like spectrum, and for which we expect to be
able to match gauge theory and string theory. In particular, we
expect to find semi-classical string states in planar $\CN=4$ SYM on
$\R \times S^3$ in the regime \eqref{stringregime}.

Note that it is clear from \eqref{stringregime} that the $|D-J| \ll
\lambda \ll 1$ requirement means that only states with $D_0=J$ can
be present. Thus, \eqref{stringregime} is a alternative way of
representing the perhaps most interesting part of our decoupled theories without resorting to limits.

As we comment on further in the Conclusions in Section
\ref{sec:concl}, it would be highly interesting to examine the
regimes \eqref{stringregime} of $\CN=4$ SYM on $\R \times S^3$ further. In these regimes
one can hope to find precise matches between weakly coupled gauge theory and
string theory.

Finally, it is important to explain why we only consider the nine
theories with scalars, and not the fermionic $SU(1,1)$, the
$SU(1,2)$ and the $SU(1,2|1)$ theories. This is due to the presence
of the $cL$ term in the dispersion relation
\eqref{eq:genBetheSpectrum}, with non-zero $c$. This means that for
large $\tilde{\lambda}$ there is a $\tilde{\lambda} c L$ term in
$\tilde{\lambda} D_2$. With such a term one cannot connect having
large $\tilde{\lambda}$ to the low energy behavior of $D_2$. Hence
the regime \eqref{stringregime} does not apply for these three
theories. This is another manifestation of the fact that the free
limit of the spin chains and the large $\tilde{\lambda}$ limit are
not connected for these three theories, as already discussed for the
large $\tilde{\lambda}$ limit of the $SU(1,2)$ theory in Section
\ref{sec:largelambsu12}.

\section{A decoupling limit of pure Yang-Mills theory}
\label{sec:pureYM}

\newcommand{\DYM}{D^{\rm (YM)}}

In this section we consider a new decoupling limit of pure
Yang-Mills theory (YM) on $\R \times S^3$. In the planar limit, the
pure YM theory reduces in the decoupling limit to a fully integrable spin
chain. The limit is analogous to the $SU(1,2)$ limit of $\CN=4$ SYM
on $\R \times S^3$ as found in Section \ref{sec:newlim}. We
furthermore write down a microcanonical version of the limit which
also applies to gauge-invariant operators of pure YM on $\R^4$.

The pure YM Lagrangian is invariant classically under conformal
transformations. Thus, it has the conformal group in four dimensions
$SO(2,4) \simeq SU(2,2)$ as symmetry group. However, contrary to
$\CN=4$ SYM, pure YM is not a conformal theory since the conformal
symmetry is broken by quantum corrections. Specifically, the beta
function $\beta(\lambda)$ for the 't Hooft coupling of pure YM
becomes non-zero at second order in the 't Hooft coupling
$\lambda = \gym^2 N/(4\pi^2)$ \cite{PhysRevLett.30.1343,PhysRevLett.30.1346}.

Nevertheless, since the beta function is non-zero only at 2-loop
order, we can regard pure YM as being a conformal theory when
considering only the tree-level and one-loop diagrams. And this will
be enough to formulate a decoupling limit for pure YM, based on the
same considerations as for $\CN=4$ SYM.

The Cartan generators of the conformal group $SO(2,4)$ are the
dilatation operator $\DYM$ and the two Cartan generators $S_1$ and
$S_2$ for the $SO(4)$ subgroup. For small 't Hooft coupling we can
expand the dilatation operator as $\DYM = \DYM_0 + \lambda \DYM_2 +
\CO(\lambda^2)$ where $\DYM_0$ is the bare scaling dimension and
$\DYM_2$ gives the one-loop anomalous dimension (computed in
\cite{Beisert:2004fv}). We write the temperature as $T = 1/\beta$
and the chemical potentials corresponding to $S_1$ and $S_2$ as
$\omega_1$ and $\omega_2$.

Since pure YM is conformally invariant to one-loop order we can
employ the state/operator correspondence relating states of pure YM
on $\R \times S^3$ to gauge-invariant operators of pure YM on
$\R^4$. The set of gauge-invariant operators of pure YM consists of
the linear combinations of multi-trace operators that can be
constructed using the set of letters consisting of the 6 gauge field
strength components and the descendants obtained by applying the
covariant derivative. The gauge field strength and covariant
derivative components transform as in $\CN=4$ SYM, thus one can use
Tables \ref{tab:gfield} and \ref{tab:deriv} also for pure YM, if one
ignores the $SU(4)$ part.

In the following we take the two chemical potentials to be equal
$\omega_1=\omega_2=\omega$ and consider the decoupling limit of
pure YM with gauge group $SU(N)$ on $\R \times S^3$ given by
\begin{equation}
\label{pureYMlim} \beta \rightarrow \infty, \quad
\tilde{\beta} \equiv \beta (1-\omega) \ \mbox{fixed}, \quad
\tilde{\lambda} \equiv \frac{\lambda}{1-\omega} \ \mbox{fixed}, \quad
N \ \mbox{fixed}.
\end{equation}
By the same arguments as in Section \ref{sec:gencon}, one sees that
the complete partition function in the grand canonical ensemble of
pure YM with gauge group $SU(N)$ on $\R \times S^3$ in the limit
\eqref{pureYMlim} reduces to
\begin{equation}
\label{partYM} Z_{\tilde{\lambda},N}(\tilde{\beta})  = {\tr}_\CH
\left[ \exp \left\{ - \tilde{\beta} \left( \DYM_0 + \tilde{\lambda}
\DYM_2 \right) \right\} \right]
\end{equation}
where $\CH$ is the set of gauge-invariant operators (or the
corresponding states) obeying $\DYM_0 = S_1 + S_2$. Thus, $\CH$
consists of any linear combination of multi-trace
operators that can be written using the letters $d_1^m d_2^k
\bar{F}_+$. This set of letters transforms in the $[0,-3]$
representation of the $su(1,2)$ algebra, hence the decoupled theory
has a $SU(1,2)$ symmetry. This can be seen by employing the same
arguments as for the $SU(1,2)$ limit of $\CN=4$ SYM on $\R \times
S^3$.

We thus see that in the decoupling limit \eqref{pureYMlim} only the
operators in $\CH$, made from the letters $d_1^m d_2^k \bar{F}_+$,
contribute to the partition function. All the other gauge-invariant
operators of pure YM are decoupled. Moreover, we have an effective
Hamiltonian $\DYM_0 + \tilde{\lambda} \DYM_2$. From
\cite{Beisert:2004fv} we have that $\DYM_2 = D_2 - \frac{11}{12} L$ for
operators in $\CH$, where $D_2$ is the truncation of the one-loop
contribution to the dilatation operator of $\CN=4$ SYM in the
$SU(1,2)$ decoupled theory. Using this, we can translate the results for the
$SU(1,2)$ theory in $\CN=4$ SYM to pure YM in the decoupling limit
\eqref{pureYMlim}.

We now turn to the planar limit of pure YM on $\R \times S^3$. Here
we can focus on single-trace operators, and they can be interpreted
as states in a spin chain. It has been shown in
\cite{Beisert:2004fv} that planar pure YM is integrable to one loop
when restricting to chiral operators. Since the set of operators
$\CH$ is chiral, we inherit the integrability for the full chiral
sector in our decoupling limit \eqref{pureYMlim}. Moreover, since
our full Hamiltonian $\DYM_0+ \tilde{\lambda} \DYM_2$ only
contains tree-level and one-loop terms, our decoupled theory is
fully integrable.

Since the decoupled theory has a $SU(1,2)$ symmetry, the spectrum
can be found from an $SU(1,2)$ spin chain. In detail, the spectrum
follows from the dispersion relation
\begin{equation}
\DYM_2 = \frac{1}{2} \sum_{k=1}^K \frac{|V_{j_k}|}{u_k^2 +
\frac{1}{4}V_{j_k}^2} + \frac{7}{12} L
\end{equation}
along with the Bethe equation \eqref{eq:genBethe1}, inserting here
that we are in the $[0,-3]$ representation of $su(1,2)$, and the
cyclicity condition \eqref{eq:genBethe2} with $U=1$. This gives the
full spectrum of pure YM on $\R\times S^3$ in the decoupling limit
\eqref{pureYMlim}.

We can furthermore follow our computations of Section \ref{sec:temp}
and obtain the thermodynamics of the decoupled theory. First, the
letter partition function for pure YM is given by \eqref{vector}.
Taking the limit \eqref{pureYMlim} of this we get
\begin{equation}
\label{pureYMlettpart} z(\tilde{x}) =
\frac{\tilde{x}^2}{(1-\tilde{x})^2}
\end{equation}
as for the $SU(1,2)$ limit of $\CN=4$ SYM on $\R \times S^3$.
Computing furthermore the expectation value of $\DYM_2$, we get
\begin{equation}
\label{pureYMexpD2} \langle \DYM_2(\tilde x)\rangle=
\frac{\tilde{x}^2 \left[ \left(1- \frac{11}{12} \tilde{x}\right)
\tilde{x} + (1-2\tilde{x}) \log ( 1 - \tilde{x}) \right] }
{(1-\tilde{x})^4}.
\end{equation}
Using \eqref{pureYMlettpart} and \eqref{pureYMexpD2} along with
\eqref{1loop} we get the Hagedorn temperature in the decoupled
theory to first order in $\tilde{\lambda}$
\begin{equation}
\tilde{T}_\textrm{H} = \frac{1}{\log 2} + \tilde{\lambda} \frac{13}{48\log 2}
+ \CO(\tilde{\lambda}^2 ).
\end{equation}

Turning instead to large $\tilde{\lambda}$, we run into
the same difficulties as encountered in Section
\ref{sec:largelambsu12} for the $SU(1,2)$ decoupling limit of
$\CN=4$ SYM on $\R \times S^3$. Defining the function
\begin{equation}
V_{\rm (YM)} (\tilde{\beta}) \equiv \lim_{L\to\infty}
\frac{1}{L}\log\left[ {\tr}_L \left( e^{-\tilde\beta
\tilde{\lambda} \left(\DYM_2 - \frac{7}{12} L\right)}\right) \right]
\end{equation}
we see that in general the Hagedorn temperature satisfies the
equation
\begin{equation}
\label{hej}
\left(2 + \frac{7}{12} \tilde\lambda\right) \tilde{\beta}_\textrm{H}
= V_{\rm (YM)} (\tilde{\beta}_\textrm{H}).
\end{equation}
However, while the left-hand side of \eqref{hej} goes to infinity as
$\tilde{\lambda} \tilde{\beta}_\textrm{H} \rightarrow \infty$, the right-hand
side goes to zero, in parallel with the analysis of Section
\ref{sec:largelambsu12}. Thus, we cannot use the free limit of
the spin chain to infer the large $\tilde{\lambda}$ behavior of the
Hagedorn temperature. Following the discussion in Section
\ref{sec:largelambsu12}, this has the consequence that a string dual
of pure YM on $\R \times S^3$ in the decoupling limit
\eqref{pureYMlim} will be difficult to find, since one cannot
consider a limit wherein the world-sheet theory of the strings is
free. More generally, this suggests that a string dual of pure YM
will be difficult to attain.

As for the decoupling limits of $\CN=4$ SYM, we can write the
decoupling limit \eqref{pureYMlim} as a decoupling limit in the
microcanonical ensemble, following Section \ref{sec:micro}. This
microcanonical decoupling limit of pure YM with gauge group $SU(N)$
takes the form
\begin{equation}
\label{YMmiclim} \lambda \rightarrow 0, \quad
\tilde{H} \equiv \frac{\DYM-S_1-S_2}{\lambda} \ \mbox{fixed}, \quad
S_1+S_2,\, N \ \mbox{fixed}.
\end{equation}
This limit can also  be thought of as a decoupling limit for
gauge-invariant operators of pure YM on $\R^4$.

The search for integrable structures in pure YM and QCD has received
considerable attention recently \cite{Ferretti:2004ba,Beisert:2004fv}.
In \cite{Beisert:2004fv}
the full one-loop anomalous dimension matrix has been computed and
studied, finding a large integrable structure in the chiral sectors.
The decoupling limit \eqref{pureYMlim} gives a decoupled
sector which is a subsector of one of the chiral sectors. However, the advantage of our decoupling limit
\eqref{pureYMlim} is that after the limit we get a decoupled theory
which is fully integrable. This enables us to study what happens in
a strong coupling limit, which for the decoupled theory is
$\tilde{\lambda} \rightarrow \infty$.

Finally, we remark that it was conjectured in \cite{Aharony:2003sx}
that the Hagedorn phase transition in weakly coupled pure YM on $\R
\times S^3$ is continuously connected to the
confinement/deconfinement transition in pure YM on $\R^4$. This
suggest that our above results perhaps can be useful to learn more
about the confinement/deconfiment transition in pure YM.


\section{Discussion and conclusions}
\label{sec:concl}

The general idea of this paper is to consider $\CN=4$ SYM on $\R
\times S^3$ near critical points with zero temperature and critical
chemical potentials. Analyzing $\CN=4$ SYM on $\R \times S^3$ in
such a near-critical region gives rise to fourteen different
decoupled theories that are a good description of weakly coupled
$\CN=4$ SYM on $\R \times S^3$ near fourteen different critical
points. The precise formulation of this is in terms of the
decoupling limits \eqref{declim} which are taken of the partition
function in the grand canonical ensemble. Taking these limits we
decouple physically fourteen different theories contained in $\CN=4$
SYM that are much simpler than the full theory but still share many
of its interesting features. The chemical potentials that we have
are the two chemical potentials for the $SO(4)$ symmetry and the
three chemical potentials for the $SU(4)$ R-symmetry. The analysis
of the near-critical regions generalizes the one of
\cite{Harmark:2006di} where only the R-symmetry chemical potentials
were considered.

For each of the decoupled theories we found an effective Hamiltonian
of the form $D_0+ \tilde{\lambda}D_2$. This Hamiltonian is valid for
any value of $\tilde{\lambda}$, thus we can study the decoupled
theory for any value of $\tilde{\lambda}$ since both $D_0$ and $D_2$
are known explicitly. We used this fact to study the planar limit,
where for each of the fourteen theories $D_2$ is equivalent to a
Hamiltonian for an integrable spin chain. In the theories with
scalars we used this to determine the spectrum and the Hagedorn
temperature in the limit of large $\tilde{\lambda}$. In this sense
we see that we are able to take explicitly a strong coupling limit
for these nine decoupled theories.

One of the decoupling limits gives rise to a decoupled theory with
$SU(1,2|3)$ symmetry. We have shown that this particular theory
contains all of the other thirteen decoupled theories. Note that
this theory also contains the half-BPS operators of $\CN=4$ SYM
since they all satisfy the relation $D = S_1+S_2+J_1+J_2+J_3$.

The $SU(1|1)$ decoupled theory is particularly interesting since in
the planar limit it corresponds to an exactly solvable spin chain,
namely the Heisenberg $XX_{1/2}$ spin chain coupled to an external
magnetic field. Thus, for this decoupled theory the exact partition
function can be found. Using this we obtained an exact equation that
determines the Hagedorn temperature as a function of
$\tilde{\lambda}$, from which the small and large $\tilde{\lambda}$
expansions are easily infered.

Another interesting decoupled theory that we studied is the one with
$SU(1,2)$ symmetry. For this theory, it is considerably harder to
take the large $\tilde{\lambda}$ limit. This is seen by considering
the planar limit, for which we find that in the free magnon spectrum
of $D_2$ the ground state energy is moved up from zero to a value
proportional to the length of the spin chain $L$, contrary to what
happens for the nine non-trivial theories with scalars.

The pure YM decoupling limit \eqref{pureYMlim} gives rise to a
decoupled theory which is almost identical to the $SU(1,2)$
decoupled theory of $\CN=4$ SYM. This is interesting in view of the
problems with taking the large $\tilde{\lambda}$ limit since they
translate to the pure YM decoupled theory. This suggests that it is
hard to find a string-dual of pure YM, since our results imply that
one cannot find a regime in which the world-sheet theory is free.

We identified an equivalent formulation of the decoupling limits in
terms of the microcanonical ensemble. This is important since it
gives a better understanding of which regime of the theory we
zoom in to when going near one of the critical points. We used in
particular these insights to determine the regimes
\eqref{stringregime} of $\CN=4$ SYM in which we have string-like
states.

\subsubsection*{Future directions and outlook}

Inspired by the work \cite{Harmark:2006ta,Harmark:2006ie}, one of
the interesting future directions is to find the decoupling limits
for type IIB strings on $\ads_5\times S^5$ that are dual to the
gauge theory decoupling limits found in this paper
\cite{Workinprogress}. We expect this to be possible for the nine
decoupling limits for which the decoupled theories have scalars. It
would in particular be interesting to find Penrose limits consistent
with the decoupling limits, enabling one to match the spectra on the
gauge and string sides in the large $\tilde{\lambda}$ limit and for
long operators.

Following \cite{Harmark:2006ie} it would be interesting to examine
the more general decoupling limits for which one obtains effective
chemical potentials in the decoupled theories. For example for the
$SU(1,2|3)$ limit one has four effective chemical potentials coming
from the differences $\omega_1-\omega_2$, $\omega_1-\Omega_1$,
$\Omega_1-\Omega_2$ and $\Omega_2-\Omega_3$. These four effective
chemical potentials should then correspond to turning on four
magnetic fields in the $SU(1,2|3)$ spin chain, and furthermore
correspond to having four rotation angles on the dual pp-wave
background.

A particularly important aspect of the decoupling limit
\eqref{declim} is that it could allow to directly investigate the
validity of the AdS/CFT correspondence. This is realized by the fact
that on the gauge theory side we can take a strong $\tilde \lambda$
limit even though the 't Hooft coupling $\lambda$ goes to zero in
the limit \eqref{declim}. The strong $\tilde\lambda$ regime should
then be related via the AdS/CFT correspondence to the string theory
dual of the gauge theory under investigation. This means that we can
study weakly coupled gauge theory and string theory in the same
regime and thus we can hope to compare the computations on both
sides directly.

As explained in the main text, the decoupling limit \eqref{declim}
is defined also for finite values of $N$, $N$ being the number of colors. Thus, using the decoupling
limit \eqref{declim} one can obtain a very convenient environment
where to compute the non-planar corrections to the gauge theory
partition function. We expect that this will allow one to gain more
information about important aspects of the Hagedorn/deconfinement
phase transition. For example it should then be possible to study
interesting questions such as what the order of the phase transition
is or what the nature of the phase above the Hagedorn transition is,
and one could furthermore hope to understand the behavior of the
theory for very high temperatures both at weak and strong coupling
$\tilde{\lambda}$. Employing the fact that our decoupling limits work for finite $N$ we can also hope to understand effects for black holes in $\ads_5\times S^5$. This could potentially lead to a better understanding of such important issues as the unitarity of black hole physics and the microstates of black holes.

Finally, it would be interesting to generalize our results to other
gauge theories. In particular, it would be interesting to study
decoupling limits of thermal $\CN=4$ SYM on $\R \times S^3 / \Z_k$
and of the dimensionally reduced 2+1 dimensional SYM theory on $\R
\times S^2$
\cite{Lin:2005nh,Ishiki:2006rt,Hikida:2006qb,Papadodimas:2006jd,Grignani:2007xz}.
This would be interesting in view of the effects of the non-trivial
vacua, and here a decoupling limit of the kind presented in this
paper could be essential to study the theories beyond the
zero-coupling regime.


\section*{Acknowledgments}

We thank the Galileo Galilei Institute for Theoretical Physics for
the hospitality and the INFN for partial support during the
completion of this work.
T.H. thanks the Carlsberg Foundation for support. The work of M.O. is supported
in part by the European Community's Human Potential Programme under
contract MRTN-CT-2004-005104 `Constituents, fundamental forces and
symmetries of the universe'.

\begin{appendix}

\section{Oscillator representation of $u(2,2|4)$}
\label{app:oscrep}

The $\CN=4$ SYM theory has $PSU(2,2|4)$ as global symmetry. Since we
use the algebraic characterizations of the decoupled theories
extensively in the main text, we review in this appendix the
oscillator representation of $u(2,2|4)$
\cite{Gunaydin:1984fk,Beisert:2003jj}, which is a highly useful way
of representing both the algebra and the set of letters of $\CN=4$
SYM. In Appendix \ref{app:algdecsec} we use this to understand the
algebra and representation for each decoupled theory.

\subsubsection*{The generators of $u(2,2|4)$}

In the oscillator representation of $u(2,2|4)$  we consider two
bosonic oscillators $\mathbf{a}^{\alpha}$,
$\mathbf{b}^{\dot\alpha}$, $\alpha,\dot\alpha=1,2$, and one
fermionic oscillator $\mathbf{c}^{a}$, $a=1,2,3,4$, with the
commutation relations \cite{Gunaydin:1984fk,Beisert:2003jj}
\begin{equation}
\label{osccom}
\left[\mathbf{a}^{\alpha},{\mathbf{a}}^{\dagger}_{\beta}\right]=
\delta^{\alpha}_{\beta}~, \quad
\left[\mathbf{b}^{\dot\alpha},{\mathbf{b}}^{\dagger}_{\dot\beta}\right]
=\delta^{\dot\alpha}_{\dot\beta}~,\quad
\left\{\mathbf{c}^{a},{\mathbf{c}}^{\dagger}_{b}\right\}=\delta^a_b~.
\end{equation}
Define furthermore the number operators
\begin{equation}
a^\alpha = {\mathbf{a}}^{\dagger}_{\alpha} \mathbf{a}^{\alpha} \spa
b^{\dot\alpha} = {\mathbf{b}}^{\dagger}_{\dot\alpha}
\mathbf{b}^{\dot\alpha} \spa c^a = {\mathbf{c}}^{\dagger}_{a}
\mathbf{c}^{a}
\end{equation}
where we do not sum over the indices, and we have
$\alpha,\dot\alpha=1,2$ and $a=1,2,3,4$.

In terms of the oscillators the $so(4) \simeq su(2) \times su(2)$
subalgebra of $u(2,2|4)$ has the 6 generators given by
\begin{equation}
{\mathbf{L}^\alpha}_\beta = {\mathbf{a}}^{\dagger}_{\beta}
\mathbf{a}^{\alpha} - \frac{a^1+a^2}{2} \delta^\alpha_\beta
 \spa
{\dot{\mathbf{L}}}^{\dot{\alpha}} {}_{\dot{\beta}} =
{\mathbf{b}}^{\dagger}_{\dot\beta} \mathbf{b}^{\dot\alpha} -
\frac{b^1+b^2}{2} \delta^{\dot\alpha}_{\dot\beta}\ .
\end{equation}
The 15 generators of the $su(4)$ subalgebra are
\begin{equation}
\label{su4alg} {\mathbf{R}^a}_b = {\mathbf{c}}^{\dagger}_{b}
\mathbf{c}^{a} - \frac{1}{4} \delta^a_b \sum_{d=1}^4 c^d ~.
\end{equation}
We have three $u(1)$ charges being the bare dilatation operator%
\footnote{Note that here we are concerned with the $psu(2,2|4)$
algebra of $\CN=4$ SYM for zero gauge coupling, $i.e.$
$\lambda=0$.} $D_0$, the central charge $C$ and the hypercharge $B$,
given as
\begin{eqnarray}
\label{theu1s} D_0 &=& 1 + \frac{1}{2} (a^1+a^2+b^1+b^2)
\nn \\
C &=& 1 + \frac{1}{2} ( - a^1-a^2+b^1+b^2-c^1-c^2-c^3-c^4)
\nn \\
B &=& \frac{1}{2} (a^1+a^2-b^1-b^2).
\end{eqnarray}
In addition to this, we have four translation generators $\mathbf{P}_{\alpha
\dot{\beta}}$ and four boost generators
$\mathbf{K}^{\alpha\dot{\beta}}$ given by
\begin{equation}
\label{PKalg}
\mathbf{P}_{\alpha \dot{\beta}} = {\mathbf{a}}^{\dagger}_{\alpha}
{\mathbf{b}}^{\dagger}_{\dot\beta} \spa
\mathbf{K}^{\alpha\dot{\beta}} = \mathbf{a}^{\alpha}
\mathbf{b}^{\dot\beta}
\end{equation}
and the 32 fermionic generators
\begin{equation}
\label{susyalg}
\mathbf{Q}^a {}_\alpha = {\mathbf{a}}^{\dagger}_{\alpha}
\mathbf{c}^{a}  \spa \dot{\mathbf{Q}}_{\dot{\alpha} a} =
{\mathbf{b}}^{\dagger}_{\dot\alpha} {\mathbf{c}}^{\dagger}_{a}
\spa
 \mathbf{S}^\alpha {}_a =
{\mathbf{c}}^{\dagger}_{a} \mathbf{a}^{\alpha}  \spa
\dot{\mathbf{S}}^{\dot{\alpha} a} = \mathbf{b}^{\dot\alpha}
\mathbf{c}^{a}.
\end{equation}
The set of 32 bosonic generators (${\mathbf{R}^a}_b$,
${\mathbf{L}^\alpha}_\beta$, ${\dot{\mathbf{L}}}^{\dot{\alpha}}
{}_{\dot{\beta}}$, $D_0$, $C$, $B$, $\mathbf{P}_{\alpha
\dot{\beta}}$ and $\mathbf{K}^{\alpha\dot{\beta}}$) and the 32
fermionic generators ($\mathbf{Q}^a {}_\alpha$,
$\dot{\mathbf{Q}}_{\dot{\alpha} a}$, $\mathbf{S}^\alpha {}_a$ and
$\dot{\mathbf{S}}^{\dot{\alpha} a}$) together comprise the algebra
of $u(2,2|4)$. The commutation relations can be worked out
explicitly using the commutation relations \eqref{osccom} for the
oscillators.

One can consistently drop the hypercharge $B$ from the $u(2,2|4)$
algebra, revealing $su(2,2|4)$. If one sticks to representations
with $C=0$, one can furthermore take out $C$ of the algebra, which
means going from $su(2,2|4)$ to the $psu(2,2|4)$ algebra that is the
algebra for the global symmetries of $\CN=4$ SYM.

It is useful to connect here the Cartan subalgebra of $u(2,2|4)$ in
terms of the oscillator representation to the notation that we
employ in the main text. In addition to the three $u(1)$ charges
$D_0$, $C$ and $B$ defined in \eqref{theu1s}, we have the following
Cartan generators of the $so(4) \simeq su(2) \times su(2)$  algebra
\begin{equation}
S_L = \frac{1}{2} ( a^1 - a^2 ) \spa S_R = \frac{1}{2} ( b^1 - b^2 )
\end{equation}
with the relation that $S_L = (S_1-S_2)/2$ and $S_R =(S_1+S_2)/2$,
$S_1$ and $S_2$ being the $so(4)$ Cartan generators. The $so(4)$
Cartan generators are thus
\begin{equation}
\label{Ss} S_1 = \frac{1}{2} ( a^1 - a^2 + b^1 - b^2 ) \spa S_2 =
\frac{1}{2} ( - a^1 + a^2 + b^1 - b^2 ).
\end{equation}
The Cartan generators we use for $su(4)$ are
\begin{equation}
\label{Js} J_1 = \frac{1}{2} (-c^1-c^2+c^3+c^4) \spa J_2 =
\frac{1}{2}(-c^1 + c^2 - c^3 + c^4) \spa J_3 = \frac{1}{2}(c^1 - c^2
- c^3 + c^4).
\end{equation}

Since $u(2,2|4)$ has fermionic generators it is not unique how to
split up the generators into raising and lowering operators. The
choice we use in almost all cases is the one dubbed the ``Beauty" in
\cite{Beisert:2003yb} and corresponds to choosing $\mathbf{S}^\alpha
{}_a$ and $\dot{\mathbf{S}}^{\dot{\alpha} a}$ as the fermionic
raising operators. The Dynkin diagram of the ``Beauty" is
\begin{align}
\bigcirc \!\!-\!\!\!-\!  {\textstyle \bigotimes} \!\!-\!\!\!-\!\!
\bigcirc \!\!-\!\!\!-\!\! \bigcirc \!\!-\!\!\!-\!\! \bigcirc \!\!-\!\!\!-\!\!
{\textstyle \bigotimes} \!\!-\!\!\!-\! \bigcirc \label{beauty}
\end{align}
Here the $\bigcirc$ refers to a bosonic root, while ${\textstyle
\bigotimes}$ refers to a fermionic root. We note that up to an
overall sign the Cartan matrix $M$ is uniquely determined by the
Dynkin diagram (see for example \cite{Frappat:1987ix} for the rules of
constructing the Cartan matrix). The lowering operators of
$u(2,2|4)$ corresponding to minus the simple roots associated with
the Dynkin diagram \eqref{beauty} are
\begin{equation}
\label{eq:AppC-Beauty}
{\mathbf{a}}^{\dagger}_{1} \mathbf{a}^{2} \spa
{\mathbf{a}}^{\dagger}_{2} \mathbf{c}^{1}  \spa
{\mathbf{c}}^{\dagger}_{1} \mathbf{c}^{2} \spa
{\mathbf{c}}^{\dagger}_{2} \mathbf{c}^{3} \spa
{\mathbf{c}}^{\dagger}_{3} \mathbf{c}^{4} \spa
{\mathbf{b}}^{\dagger}_{2} {\mathbf{c}}^{\dagger}_{4} \spa
{\mathbf{b}}^{\dagger}_{1} \mathbf{b}^{2}.
\end{equation}
We see that the three bosonic roots in the middle of \eqref{beauty}
correspond to the $su(4)$ R-symmetry algebra. We choose the
diagonal of the Cartan matrix to be positive for these three roots.
With this, the Cartan matrix is
\begin{align}
\label{eq:Cartanmatrix}
M= \left(\begin{array}{r|r|ccc|c|c}
-2 & +1 &&&&\\
\hline
+1 &  & -1 &&&\\
\hline
& -1 & +2 & -1 &&\\
& & -1 & +2 & -1& \\
& & & -1 & +2 & -1 \\
\hline
&&&&-1&&+1\\
\hline &&&&&+1&-2
\end{array}\right).
\end{align}
The Dynkin labels corresponding to \eqref{beauty} are
\begin{equation}
\label{beautylabels} [s_1,r_1,q_1,p,q_2,r_2,s_2]
\end{equation}
with $s_1 = a^2-a^1$ and $s_2 = b^2-b^1$ corresponding to the
$su(2)\times su(2)$ subgroups, $q_1=c^2-c^1$, $p=c^3-c^2$ and
$q_2 = c^4-c^3$ corresponding to the $su(4)$ subgroup, and finally for
the two fermionic roots we have $r_1 = a^2+c^1$ and $r_2 = 1+b^2-c^4$.

Another choice for the raising and lowering operators is the one
dubbed the ``Beast" in \cite{Beisert:2003yb} and corresponds to
choosing $\mathbf{S}^\alpha {}_a$ and
$\dot{\mathbf{Q}}_{\dot{\alpha} a}$ as the fermionic raising
operators. This choice is useful for the $SU(1,2)$ decoupled theory.
The Dynkin diagram is
\begin{equation}
\label{beast}
\bigcirc \!\!-\!\!\!-\!\! \bigcirc \!\!-\!\!\!-\!\! \bigcirc \!\!-\!\!\!-\!  {\textstyle \bigotimes} \!\!-\!\!\!-\!\!
\bigcirc \!\!-\!\!\!-\!\! \bigcirc \!\!-\!\!\!-\! \bigcirc
\end{equation}
The lowering operators corresponding to minus the simple roots associated with the Dynkin diagram are
\begin{equation}
{\mathbf{a}}^{\dagger}_{1} \mathbf{a}^{2} \spa
{\mathbf{a}}^{\dagger}_{2} {\mathbf{b}}^{\dagger}_{2} \spa
{\mathbf{b}}^{\dagger}_{1} \mathbf{b}^{2} \spa
\mathbf{b}^{1} \mathbf{c}^{1} \spa
{\mathbf{c}}^{\dagger}_{1} \mathbf{c}^{2} \spa
{\mathbf{c}}^{\dagger}_{2} \mathbf{c}^{3} \spa
{\mathbf{c}}^{\dagger}_{3} \mathbf{c}^{4} .
\end{equation}
The first three roots of \eqref{beast} correspond to the $su(2,2)$
subalgebra. We choose the Cartan matrix to be positive in the
diagonal for these nodes. The Dynkin labels for the three nodes of
$su(2,2)$ are $[s_1,r,s_2]$ with $r = - 1 -a^2-b^2$.

\subsubsection*{The letters of $\CN=4$ SYM}

In Section \ref{sec:newlim} we described the set of letters $\CA$ of
$\CN=4$ SYM. The letters are listed in Tables
\ref{tab:gfield}--\ref{tab:fermions3} along with the four components
of the covariant derivative in Table \ref{tab:deriv} using which one obtain
the descendants. In terms of the oscillators $\mathbf{a}^{\alpha}$,
$\mathbf{b}^{\dot\alpha}$, $\alpha,\dot\alpha=1,2$, and
$\mathbf{c}^{a}$, $a=1,2,3,4$, the set of letters $\CA$ of $\CN=4$
SYM is given by
\begin{equation}
\begin{array}{c}
\phi \ :\ ({\mathbf{c}}^{\dagger})^2 |0 \rangle
 \mbox{ repr. } [0,1,0]_{(0,0)} \\[1mm]
\chi \ :\ {\mathbf{a}}^{\dagger} {\mathbf{c}}^{\dagger} |0 \rangle
\mbox{ repr. } [0,0,1]_{(\frac{1}{2},0)} \spa \bar{\chi} \ :\
{\mathbf{b}}^{\dagger} ({\mathbf{c}}^{\dagger})^3 |0 \rangle
 \mbox{ repr. }  [1,0,0]_{(0,\frac{1}{2})}  \\[1mm]
F \ :\  ({\mathbf{a}}^{\dagger})^2 |0 \rangle \mbox{ repr. }
[0,0,0]_{(1,0)}  \spa \bar{F} \ :\ ({\mathbf{b}}^{\dagger})^2
({\mathbf{c}}^{\dagger})^4 |0 \rangle \mbox{ repr. }
[0,0,0]_{(0,1)}  \\[1mm]
d  \ :\ {\mathbf{a}}^{\dagger} {\mathbf{b}}^{\dagger} \mbox{ repr. }
[0,0,0]_{(\frac{1}{2},\frac{1}{2})}
\end{array}
\label{theletters}
\end{equation}
It is an easy exercise to find the explicit oscillator
representation for each letter of $\CN=4$ SYM by combining
\eqref{theletters} with the Cartan generators \eqref{Ss} and
\eqref{Js} for the $so(4)$ and $su(4)$ algebras and with the Tables
\ref{tab:gfield}--\ref{tab:deriv} of the letters.

All the letters of $\CN=4$ SYM have $C=0$ thus the set of letters $\CA$
corresponds to a representation of $psu(2,2|4)$. Considering the
``Beauty" \eqref{beauty} one can see from the hermitian conjugate of
\eqref{eq:AppC-Beauty} that the letter $Z = {\mathbf{c}}^{\dagger}_3
{\mathbf{c}}^{\dagger}_4 |0 \rangle$ is the highest weight of the
representation. Therefore, using the Dynkin labels
\eqref{beautylabels}, we see that the set of letters $\CA$ corresponds to
the $[0,0,0,1,0,0,0]$ representation of $psu(2,2|4)$. This
representation is known as the Singleton representation.

\section{Algebras and representations for decoupled theories}
\label{app:algdecsec}

We describe in this appendix briefly the algebras and
representations for each of the twelve non-trivial decoupled
theories. This is done in terms of the oscillator representation of
$u(2,2|4)$ reviewed in Appendix \ref{app:oscrep} where also the
notation used below is defined. Each decoupled theory corresponds to
a sector of $\CN=4$ SYM, $i.e.$ a subset of the full set of letters
with a sub-algebra of the full algebra $psu(2,2|4)$. The algebras
and representations can be derived from the ``Beauty" Dynkin diagram
\eqref{beauty}, except for the $SU(1,2)$ decoupled theory which is
derived from the ``Beast" Dynkin diagram \eqref{beast}.

The $SU(2)$ sector is given by
\begin{align}
c^1=0, \quad c^4=1, \quad a^1=a^2=b^1=b^2=0.
\end{align}
From the Dynkin diagram \eqref{beauty} of the full $psu(2,2|4)$
algebra we keep only the ${\mathbf{c}}^{\dagger}_{2} \mathbf{c}^{3}$
node which has Dynkin label $[p]=[c^3-c^2]$.  The highest weight
within this subsector is $Z$ which gives $[p]=[1]$.  This is twice
the spin which fits with this being the spin $1/2$ representation.

The bosonic $SU(1,1)$ sector is defined by
\begin{align}
c^1 = c^2 = 0 , \quad c^3 = c^4 = 1 , \quad a^2 = b^2 = 0.
\end{align}
The Dynkin diagram has one bosonic node with
\begin{align}
\mathbf{a}^{\dagger}_{1} \mathbf{b}^{\dagger}_{1} =
(\mathbf{a}^{\dagger}_{1} \mathbf{a}^{2})
(\mathbf{a}^{\dagger}_{2} \mathbf{b}^{\dagger}_{2})
(\mathbf{b}^{\dagger}_{1} \mathbf{b}^{2})
\end{align}
as the lowering operator.
The Dynkin label is $[r'] = [-1 - a^1 - b^1]$ which for $Z$ gives
$[-1]$.  This is again twice the spin of the representation which
fits with this being the spin $-1/2$ representation.

The fermionic $SU(1,1)$ sector is given by
\begin{align}
c^1 = c^2 = c^3 = 0 , \quad c^4 = 1 , \quad a^2 = b^2 = 0.
\end{align}
The Dynkin diagram is the same as for the bosonic $SU(1,1)$ case but
in this sector the fermion $\chi_1$ is the highest weight and the
Dynkin label becomes $[r'] =  [-2]$ which fits well with this being
the spin $-1$ representation.

The $SU(1|1)$ sector is given by
\begin{align}
c^1 = c^2 = 0 , \quad c^4 = 1 , \quad a^2 = b^1 = b^2 = 0.
\end{align}
The Dynkin diagram has one fermionic node with
\begin{align}
\mathbf{a}^{\dagger}_{1} \mathbf{c}^{3} =
({\mathbf{a}}^{\dagger}_{1} \mathbf{a}^{2})
({\mathbf{a}}^{\dagger}_{2} \mathbf{c}^{1})
({\mathbf{c}}^{\dagger}_{1} \mathbf{c}^{2})
({\mathbf{c}}^{\dagger}_{2} \mathbf{c}^{3})
\end{align}
as the simple lowering operator. The Dynkin label is
$[r_1'] = [a^1 + c^3]$ and for the highest weight $Z$ we get
$[r_1'] = [1]$.

The $SU(1|2)$ sector is defined by
\begin{align}
c^1 = 0 , \quad c^4 = 1 , \quad a^1 = a^2 = b^2 = 0.
\end{align}
This is our first example of a symmetry algebra with rank higher
than one. The Dynkin diagram has in this case two nodes, one bosonic
and one fermionic
\begin{align}
\bigcirc \!\!-\!\!\!-\!\! \textstyle \bigotimes
\end{align}
The lowering operators corresponding to these nodes are
\begin{align}
{\mathbf{c}}^{\dagger}_{2} \mathbf{c}^{3}, \quad
{\mathbf{b}}^{\dagger}_{1} \mathbf{c}^\dagger_{3}=
({\mathbf{c}}^{\dagger}_{3} \mathbf{c}^{4})
({\mathbf{b}}^{\dagger}_{2} {\mathbf{c}}^{\dagger}_{4})
({\mathbf{b}}^{\dagger}_{1} \mathbf{b}^{2}).
\end{align}
The Dynkin labels are $[p,r_2'] = [c^3-c^2, 1+b^1-c^3]$ which
for the highest weight $Z$ gives $[1,0]$.

The $SU(2|3)$ sector is defined by
\begin{align}
c^4 = 1 , \quad b^1 = b^2 = 0.
\end{align}
For this sector we keep the first four nodes of the
Beauty diagram
\begin{align}
\bigcirc \!\!-\!\!\!-\!\! \textstyle \bigotimes  \!\!-\!\!\!-\!\! \bigcirc \!\!-\!\!\!-\! \bigcirc
\end{align}
with lowering operators
\begin{equation}
{\mathbf{a}}^{\dagger}_{1} \mathbf{a}^{2} ,\quad
{\mathbf{a}}^{\dagger}_{2} \mathbf{c}^{1} ,\quad
{\mathbf{c}}^{\dagger}_{1} \mathbf{c}^{2} ,\quad
{\mathbf{c}}^{\dagger}_{2} \mathbf{c}^{3} .
\end{equation}
The Dynkin labels for this sector are $[s_1,r_1,q_1,p]$ which
for the highest weight $Z$ gives $[0,0,0,1]$.

The $SU(1,1|1)$ sector is defined by
\begin{align}
c^1 = 0 , \quad c^3 = c^4 = 1 , \quad a^2 = b^2 = 0.
\end{align}
We can obtain all these states from the highest weight $Z$
by combining the first three lowering operators of the
Beauty into one fermionic operator
\begin{equation}
{\mathbf{a}}^{\dagger}_{1} \mathbf{c}^{2}=
({\mathbf{a}}^{\dagger}_{1} \mathbf{a}^{2})
({\mathbf{a}}^{\dagger}_{2} \mathbf{c}^{1})
({\mathbf{c}}^{\dagger}_{1} \mathbf{c}^{2})
\end{equation}
and by combining the last four lowering operators into another
fermionic operator
\begin{equation}
{\mathbf{b}}^{\dagger}_{1} {\mathbf{c}}^{\dagger}_{2} =
({\mathbf{c}}^{\dagger}_{2} \mathbf{c}^{3})
({\mathbf{c}}^{\dagger}_{3} \mathbf{c}^{4} )
({\mathbf{b}}^{\dagger}_{2} {\mathbf{c}}^{\dagger}_{4})
({\mathbf{b}}^{\dagger}_{1} \mathbf{b}^{2}).
\end{equation}
Since this sector has two fermionic roots the Dynkin diagram is
\begin{align}
\textstyle \bigotimes \!\!-\!\!\!-\!\! \textstyle \bigotimes
\end{align}
and the Dynkin labels are $[r_1',r_2']=[a^1+c^2,1+b^1-c^2]$
which for the highest weight $Z$ gives $[0,1]$.

The $SU(1,1|2)$ sector is defined by
\begin{align}
c^1 = 0 , \quad c^4 = 1 , \quad a^2 = b^2 = 0.
\end{align}
We use a similar combination of roots as in the previous sector,
except now we keep the middle operator as it is in the Beauty.
The three lowering operators that we have at our disposal are then
\begin{align}
{\mathbf{a}}^{\dagger}_{1} \mathbf{c}^{2}=
({\mathbf{a}}^{\dagger}_{1} \mathbf{a}^{2})
({\mathbf{a}}^{\dagger}_{2} \mathbf{c}^{1})
({\mathbf{c}}^{\dagger}_{1} \mathbf{c}^{2}), \qquad
{\mathbf{c}}^{\dagger}_{2} \mathbf{c}^{3}, \qquad
{\mathbf{b}}^{\dagger}_{1} {\mathbf{c}}^{\dagger}_{3} =
({\mathbf{c}}^{\dagger}_{3} \mathbf{c}^{4} )
({\mathbf{b}}^{\dagger}_{2} {\mathbf{c}}^{\dagger}_{4})
({\mathbf{b}}^{\dagger}_{1} \mathbf{b}^{2}).
\end{align}
The Dynkin diagram is
\begin{align}
\textstyle \bigotimes \!\!-\!\!\!-\!\! \bigcirc \!\!-\!\!\!-\!\! \textstyle \bigotimes
\end{align}
and the Dynkin labels are
$[r_1',p,r_2'] = [a^1+c^2, c^3-c^2, 1+b^1-c^3] = [0,1,0]$

The $SU(1,2)$ sector is defined by
\begin{align}
c^1 = c^2 = c^3 = c^4 = 1 , \quad b^2 = 0.
\end{align}
As already mentioned, we obtain this sector from the ``Beast" Dynkin
diagram \eqref{beast}. Specifically, we consider the first three
nodes of \eqref{beast} corresponding to the $su(2,2)$ algebra. We
keep the first node as it is but combine the latter two. The
lowering operators that we get in this way are
\begin{align}
{\mathbf{a}}^{\dagger}_{1} \mathbf{a}^{2}, \qquad
{\mathbf{a}}^{\dagger}_{2} {\mathbf{b}}^{\dagger}_{1} =
({\mathbf{a}}^{\dagger}_{2} {\mathbf{b}}^{\dagger}_{2})
({\mathbf{b}}^{\dagger}_{1} \mathbf{b}^{2}).
\end{align}
The Dynkin diagram is
\begin{align}
\bigcirc \!\!-\!\!\!-\! \bigcirc
\end{align}
and the Dynkin labels are $[s_1,r'] = [a^2-a^1, -1 - a^2 - b^1] = [0,-3]$.

The $SU(1,2|1)$ sector is defined by
\begin{align}
c^2 = c^3 = c^4 = 1 , \quad b^2 = 0.
\end{align}
We need three lowering operators
\begin{align}
{\mathbf{a}}^{\dagger}_{1} \mathbf{a}^{2}, \quad
{\mathbf{a}}^{\dagger}_{2} \mathbf{c}^{1}, \quad
{\mathbf{b}}^{\dagger}_{1} {\mathbf{c}}^{\dagger}_{1} =
({\mathbf{c}}^{\dagger}_{1} \mathbf{c}^{2} )
({\mathbf{c}}^{\dagger}_{2} \mathbf{c}^{3} )
({\mathbf{c}}^{\dagger}_{3} \mathbf{c}^{4} )
({\mathbf{b}}^{\dagger}_{2} {\mathbf{c}}^{\dagger}_{4})
({\mathbf{b}}^{\dagger}_{1} \mathbf{b}^{2}).
\end{align}
The Dynkin diagram is
\begin{align}
\bigcirc \!\!-\!\!\!-\!\!  \textstyle \bigotimes \!\!-\!\!\!-\!\! \textstyle \bigotimes
\end{align}
and the Dynkin labels are
$[s_1,r_1,r_2'] = [a^2-a^1, a^2+c^1, 1+ b^1-c^1] = [0,0,2]$.

The $SU(1,2|2)$ sector is given by
\begin{align}
c^3 = c^4 = 1 , \quad b^2 = 0.
\end{align}
We need four lowering operators
\begin{equation}
{\mathbf{a}}^{\dagger}_{1} \mathbf{a}^{2}, \qquad
{\mathbf{a}}^{\dagger}_{2} \mathbf{c}^{1}, \qquad
{\mathbf{c}}^{\dagger}_{1} \mathbf{c}^{2}, \qquad
{\mathbf{b}}^{\dagger}_{1} {\mathbf{c}}^{\dagger}_{2} =
({\mathbf{c}}^{\dagger}_{2} \mathbf{c}^{3})
({\mathbf{c}}^{\dagger}_{3} \mathbf{c}^{4})
({\mathbf{b}}^{\dagger}_{2} {\mathbf{c}}^{\dagger}_{4})
({\mathbf{b}}^{\dagger}_{1} \mathbf{b}^{2}).
\end{equation}
Starting from $Z$ we can get all the other allowed letters in this
sector by applying these four lowering operators in a particular
order. First let us see how to go from $Z$ to any of the other types
of fields:
\begin{align}
Z \ \to\
{\mathbf{b}}^{\dagger}_{1} {\mathbf{c}}^{\dagger}_{2} \ Z = \bchi_7  \ \to\
{\mathbf{c}}^{\dagger}_{1} \mathbf{c}^{2} \ \bchi_7 = \bchi_5 \ \to\
{\mathbf{b}}^{\dagger}_{1} {\mathbf{c}}^{\dagger}_{2} \ \bchi_5 = \bar F_+ \,.
\end{align}
We can also obtain $d_1 Z$ or $d_2 Z$ from plain $Z$:
\begin{align}
Z \ \to\
{\mathbf{b}}^{\dagger}_{1} {\mathbf{c}}^{\dagger}_{2} \ Z = \bchi_7  \ \to\
{\mathbf{c}}^{\dagger}_{1} \mathbf{c}^{2} \ \bchi_7 = \bchi_5 \ \to\
{\mathbf{a}}^{\dagger}_{2} \mathbf{c}^{1}\  \bchi_5 = d_2 Z \ \to \
{\mathbf{a}}^{\dagger}_{1} \mathbf{a}^{2}\  d_2 Z = d_1 Z.
\end{align}
Using the second chain repeatedly we can clearly get $d_1^k d_2^\ell Z$
and using the first chain we can map $d_1^k d_2^\ell Z$ to any of the
other letters with the same set of derivatives.
The Dynkin diagram is
\begin{align}
\bigcirc \!\!-\!\!\!-\!\!  \textstyle \bigotimes \!\!-\!\!\!-\!\!
\bigcirc \!\!-\!\!\!-\!\! \textstyle \bigotimes
\end{align}
and the Dynkin labels are $[s_1,r_1,q_1,r_2']$ with
$r_2' = 1+b^1 - c^2$. The highest weight is $Z$ and we get $[0,0,0,1]$.

The $SU(1,2|3)$ sector is defined by
\begin{align}
c^4 = 1 , \quad b^2 = 0.
\end{align}
We need five lowering operators and we get them from the Beauty by
combining the last three roots into one fermionic root. The
corresponding operator will be $ {\mathbf{b}}^{\dagger}_{1}
{\mathbf{c}}^{\dagger}_{3} = ({\mathbf{c}}^{\dagger}_{3}
\mathbf{c}^{4}) ({\mathbf{b}}^{\dagger}_{2}
{\mathbf{c}}^{\dagger}_{4}) ({\mathbf{b}}^{\dagger}_{1}
\mathbf{b}^{2})$ with Dynkin label $r_2' = 1+b^1-c^3$.
The highest weight is $Z$ and the Dynkin labels are
$[s_1,r_1,q_1,p,r_2'] = [0,0,0,1,0]$.
The Dynkin diagram is
\begin{align}
\bigcirc \!\!-\!\!\!-\!\!  \textstyle \bigotimes \!\!-\!\!\!-\!\!
\bigcirc \!\!-\!\!\!-\!\! \bigcirc \!\!-\!\!\!-\!\! \textstyle \bigotimes
\end{align}

The Cartan matrices for all our decoupled theories can be obtained
from the Cartan matrix of the ``Beauty" in
Eq.~\eqref{eq:Cartanmatrix} by deleting appropriate columns and rows
in accordance with the Dynkin diagram of each sector.  A fermionic
node always gives rise to a zero on the diagonal while bosonic roots
give either plus or minus two. This is with the notable exception of
the $SU(1,2)$ theory for which we have the usual Cartan matrix of
the $sl(3)$ algebra.


\section{The letter partition function}
\label{sec:freetheory2}

The result \eqref{letter} for the letter partition function for
$\CN=4$ SYM on $\R \times S^3$ in the presence of non-zero chemical
potentials for the R-charges of the $SU(4)$ R-symmetry and for the
Cartan generators of the $SO(4)$ symmetry group of $S^3$ can be also
obtained using the oscillator picture. In this formalism, the
general expression for the letter partition function is given by
\begin{eqnarray}
\label{osclettpart} &&z(x,\omega_i,y_i) = {\tr}_\CA \left( x^{D_0}
y_1^{J_1} y_2^{J_2} y_3^{J_3}\rho_1^{S_1} \rho_2^{S_2} \right)
=\sum_{a^1,a^2,b^1,b^2=0}^\infty \sum_{c^1,c^2,c^3,c^4=0}^1 \delta
(C) x^{D_0} y_1^{J_1} y_2^{J_2} y_3^{J_3}\rho_1^{S_1} \rho_2^{S_2}
\cr &=& \sum_{a^1,a^2=0}^\infty \sum_{n^1=-a^1}^{\infty}
\sum_{n^2=-a^2}^{\infty}\left[Y_S\delta
\left(\frac{n^1+n^2}{2}\right)+\delta
\left(1+\frac{n^1+n^2}{2}\right)+\delta \left(\frac{n^1+n^2}{2}-1
\right)\right.\cr &+&\left.Y_1\delta
\left(\frac{1+n^1+n^2}{2}\right)+Y_2\delta
\left(\frac{n^1+n^2-1}{2}\right)\right]
x^{1+\frac{n^1+n^2}{2}+a^1+a^2} \rho_1^{\frac{n^1-n^2}{2}+a^1-a^2}
\rho_2^{\frac{n^1-n^2}{2}}
\end{eqnarray}
where $n^k=b^k-a^k$, $k=1,2$, $y_i=\exp(\beta\Omega_i)$, $i=1,2,3$,
$\rho_j=\exp(\beta\omega_j)$, $j=1,2$,
\begin{equation}
Y_S=\sum_{i=1}^3(y_i+y_i^{-1})
\end{equation}
\begin{equation}
Y_1=(y_1y_2y_3)^{1/2}+y_1^{1/2}(y_2y_3)^{-1/2}+(y_1y_3)^{-1/2}y_2^{1/2}+(y_1y_2)^{-1/2}y_3^{1/2}
\label{y1os}
\end{equation}
\begin{equation}
Y_2=(y_1y_2y_3)^{-1/2}+y_1^{-1/2}(y_2y_3)^{1/2}+(y_1y_3)^{1/2}y_2^{-1/2}+(y_1y_2)^{1/2}y_3^{-1/2}
\label{y2os}
\end{equation}
In the second line we performed the sums over the fermionic
operators $c^a$, $a=1,2,3,4$. From equation \eqref{osclettpart} it
is easy to see that, by performing all the sums, we can derive the
contribution of scalars, vectors and fermions in the various
representation separately. In more detail, the term proportional to
$\delta \left(\frac{n^1+n^2}{2}\right)$ gives the scalar partition
function \eqref{scalar}. This can be also seen from the fact that
the scalars are given acting with $({\mathbf{c}}^{\dagger})^2 $ on
the vacuum and the delta function for the central charge is given by
$\delta \left(\frac{n^1+n^2}{2}\right)$ precisely when $\sum_{a=1}^4
c^a=2$.

The term proportional to $\delta \left(1+\frac{n^1+n^2}{2}\right)$
gives the partition function for vectors in the representation
$[0,0,0]_{(1,0)}$. In fact this is the contribution when
$\sum_{a=1}^4 c^a=0$. For $\sum_{a=1}^4 c^a=4$ we get instead the
term proportional to $\delta \left(\frac{n^1+n^2}{2}-1\right)$ that
corresponds to the partition function for vectors in the
representation $[0,0,0]_{(0,1)}$. Adding together the two
contributions we get the result \eqref{vector} for the vectors
partition function.

By similar arguments, the term proportional to $\delta
\left(\frac{1+n^1+n^2}{2}\right)$ gives the partition function
\eqref{f1} for fermions in the representation $[0,0,1]_{(1/2,0)}$
and, finally, the term proportional to $\delta
\left(\frac{n^1+n^2-1}{2}\right)$ gives the partition function
\eqref{f2} for fermions in the representation $[1,0,0]_{(0,1/2)}$.

\end{appendix}



\begin{thebibliography}{10}

\bibitem{Maldacena:1997re}
J.~Maldacena, {\em Adv. Theor. Math. Phys.} {\bf 2} (1998) 231--252,
\href{http://www.arXiv.org/abs/hep-th/9711200}{{\tt hep-th/9711200}}.

\bibitem{Gubser:2002tv}
S.~S. Gubser, I.~R. Klebanov, and A.~M. Polyakov,
\href{http://arXiv.org/abs/hep-th/0204051}{{\tt hep-th/0204051}}.

\bibitem{Witten:1998qj}
E.~Witten, {\em Adv. Theor. Math. Phys.} {\bf 2} (1998) 253,
\href{http://www.arXiv.org/abs/hep-th/9802150}{{\tt hep-th/9802150}}.

\bibitem{'tHooft:1974jz}
G.~'t~Hooft, {\em Nucl. Phys.} {\bf B72} (1974) 461.

\bibitem{Minahan:2002ve}
J.~A. Minahan and K.~Zarembo, {\em JHEP} {\bf 03} (2003) 013,
\href{http://www.arXiv.org/abs/hep-th/0212208}{{\tt hep-th/0212208}}.

\bibitem{Harmark:2006di}
T.~Harmark and M.~Orselli, {\em Nucl. Phys.} {\bf B757} (2006) 117--145,
\href{http://www.arXiv.org/abs/hep-th/0605234}{{\tt hep-th/0605234}}.

\bibitem{Harmark:2006ta}
T.~Harmark and M.~Orselli, {\em Phys. Rev.} {\bf D74} (2006) 126009,
\href{http://www.arXiv.org/abs/hep-th/0608115}{{\tt hep-th/0608115}}.

\bibitem{Harmark:2006ie}
T.~Harmark, K.~R. Kristjansson, and M.~Orselli, {\em JHEP} {\bf 02} (2007) 085,
\href{http://www.arXiv.org/abs/hep-th/0611242}{{\tt hep-th/0611242}}.

\bibitem{Harmark:2007et}
T.~Harmark, K.~R. Kristjansson, and M.~Orselli, {\em Fortschr.~Phys.} {\bf 55}
  (2007), no.~5--7, 754--759,
\href{http://www.arXiv.org/abs/hep-th/0701088}{{\tt hep-th/0701088}}.

\bibitem{Bertolini:2002nr}
M.~Bertolini, J.~de~Boer, T.~Harmark, E.~Imeroni, and N.~A. Obers, {\em JHEP}
  {\bf 01} (2003) 016,
\href{http://www.arXiv.org/abs/hep-th/0209201}{{\tt hep-th/0209201}}.

\bibitem{Beisert:2004ry}
N.~Beisert, {\em Phys. Rept.} {\bf 405} (2005) 1--202,
\href{http://www.arXiv.org/abs/hep-th/0407277}{{\tt hep-th/0407277}}.

\bibitem{Beisert:2003jj}
N.~Beisert, {\em Nucl. Phys.} {\bf B676} (2004) 3--42,
\href{http://www.arXiv.org/abs/hep-th/0307015}{{\tt hep-th/0307015}}.

\bibitem{Dolan:2002zh}
F.~A. Dolan and H.~Osborn, {\em Ann. Phys.} {\bf 307} (2003) 41--89,
\href{http://www.arXiv.org/abs/hep-th/0209056}{{\tt hep-th/0209056}}.

\bibitem{Bianchi:2003wx}
M.~Bianchi, J.~F. Morales, and H.~Samtleben, {\em JHEP} {\bf 07} (2003) 062,
\href{http://www.arXiv.org/abs/hep-th/0305052}{{\tt hep-th/0305052}}.

\bibitem{Beisert:2003tq}
N.~Beisert, C.~Kristjansen, and M.~Staudacher, {\em Nucl. Phys.} {\bf B664}
  (2003) 131--184,
\href{http://www.arXiv.org/abs/hep-th/0303060}{{\tt hep-th/0303060}}.

\bibitem{Beisert:2003yb}
N.~Beisert and M.~Staudacher, {\em Nucl. Phys.} {\bf B670} (2003) 439--463,
\href{http://www.arXiv.org/abs/hep-th/0307042}{{\tt hep-th/0307042}}.

\bibitem{Beisert:2005fw}
N.~Beisert and M.~Staudacher, {\em Nucl. Phys.} {\bf B727} (2005) 1--62,
\href{http://www.arXiv.org/abs/hep-th/0504190}{{\tt hep-th/0504190}}.

\bibitem{Sundborg:1999ue}
B.~Sundborg, {\em Nucl. Phys.} {\bf B573} (2000) 349--363,
\href{http://www.arXiv.org/abs/hep-th/9908001}{{\tt hep-th/9908001}}.

\bibitem{Polyakov:2001af}
A.~M. Polyakov, {\em Int. J. Mod. Phys.} {\bf A17S1} (2002) 119--136,
\href{http://www.arXiv.org/abs/hep-th/0110196}{{\tt hep-th/0110196}}.

\bibitem{Aharony:2003sx}
O.~Aharony, J.~Marsano, S.~Minwalla, K.~Papadodimas, and M.~Van~Raamsdonk,
{\em Adv. Theor. Math. Phys.} {\bf 8} (2004) 603--696,
\href{http://www.arXiv.org/abs/hep-th/0310285}{{\tt hep-th/0310285}}.

\bibitem{Yamada:2006rx}
D.~Yamada and L.~G. Yaffe, {\em JHEP} {\bf 09} (2006) 027,
\href{http://www.arXiv.org/abs/hep-th/0602074}{{\tt hep-th/0602074}}.

\bibitem{Spradlin:2004pp}
M.~Spradlin and A.~Volovich, {\em Nucl. Phys.} {\bf B711} (2005) 199--230,
\href{http://www.arXiv.org/abs/hep-th/0408178}{{\tt hep-th/0408178}}.

\bibitem{Bianchi:2006ti}
M.~Bianchi, F.~A. Dolan, P.~J. Heslop, and H.~Osborn,
\href{http://www.arXiv.org/abs/hep-th/0609179}{{\tt hep-th/0609179}}.

\bibitem{Dolan:2007rq}
F.~A. Dolan,
\href{http://www.arXiv.org/abs/arXiv:0704.1038 [hep-th]}{{\tt arXiv:0704.1038
  [hep-th]}}.

\bibitem{Gunaydin:1984fk}
M.~Gunaydin and N.~Marcus, {\em Class. Quant. Grav.} {\bf 2} (1985)
L11.

\bibitem{Staudacher:2004tk}
M.~Staudacher, {\em JHEP} {\bf 05} (2005) 054,
\href{http://www.arXiv.org/abs/hep-th/0412188}{{\tt hep-th/0412188}}.

\bibitem{PhysRev.127.1508}
S.~Katsura, {\em Phys. Rev.} {\bf 127} (Sep, 1962) 1508--1518.

\bibitem{PhysRevLett.30.1343}
D.~J. Gross and F.~Wilczek, {\em Phys. Rev. Lett.} {\bf 30} (Jun, 1973)
  1343--1346.

\bibitem{PhysRevLett.30.1346}
H.~D. Politzer, {\em Phys. Rev. Lett.} {\bf 30} (Jun, 1973) 1346--1349.

\bibitem{Beisert:2004fv}
N.~Beisert, G.~Ferretti, R.~Heise, and K.~Zarembo, {\em Nucl. Phys.} {\bf B717}
  (2005) 137--189,
\href{http://www.arXiv.org/abs/hep-th/0412029}{{\tt hep-th/0412029}}.

\bibitem{Ferretti:2004ba}
G.~Ferretti, R.~Heise, and K.~Zarembo, {\em Phys. Rev.} {\bf D70} (2004)
  074024,
\href{http://www.arXiv.org/abs/hep-th/0404187}{{\tt hep-th/0404187}}.

\bibitem{Workinprogress}
T.~Harmark, K.~R. Kristjansson, and M.~Orselli, {\em Work in progress}.

\bibitem{Lin:2005nh}
H.~Lin and J.~M. Maldacena, {\em Phys. Rev.} {\bf D74} (2006) 084014,
\href{http://www.arXiv.org/abs/hep-th/0509235}{{\tt hep-th/0509235}}.

\bibitem{Ishiki:2006rt}
G.~Ishiki, Y.~Takayama, and A.~Tsuchiya, {\em JHEP} {\bf 10} (2006) 007,
\href{http://www.arXiv.org/abs/hep-th/0605163}{{\tt hep-th/0605163}}.

\bibitem{Hikida:2006qb}
Y.~Hikida,
\href{http://www.arXiv.org/abs/hep-th/0610119}{{\tt hep-th/0610119}}.

\bibitem{Papadodimas:2006jd}
K.~Papadodimas, H.-H. Shieh, and M.~Van~Raamsdonk, {\em JHEP} {\bf 04} (2007)
  069,
\href{http://www.arXiv.org/abs/hep-th/0612066}{{\tt hep-th/0612066}}.

\bibitem{Grignani:2007xz}
G.~Grignani, L.~Griguolo, N.~Mori, and D.~Seminara,
\href{http://www.arXiv.org/abs/arXiv:0707.0052 [hep-th]}{{\tt arXiv:0707.0052
  [hep-th]}}.

\bibitem{Frappat:1987ix}
L.~Frappat, A.~Sciarrino, and P.~Sorba, {\em Commun. Math. Phys.} {\bf 121}
  (1989)
457--500.

\end{thebibliography}

\providecommand{\href}[2]{#2}\begingroup\raggedright\endgroup

\end{document}